\documentclass{aa}

\usepackage{graphicx}
\usepackage{xcolor}
\usepackage{colortbl}
\usepackage{booktabs}
\usepackage{siunitx}
\usepackage{multirow}
\usepackage{footmisc}

\usepackage{txfonts}
\usepackage[pdfpagelabels=false]{hyperref}	
\hypersetup{colorlinks=true,linkcolor=blue,citecolor=blue,filecolor=blue,urlcolor=blue,}
\begin{document} 
   \title{Discovery of the luminous X-ray ignition eRASSt~J234402.9$-$352640\\
   Paper I: Tidal disruption event or a rapid increase in accretion in an active galactic nucleus?
   }
   \titlerunning{eRASSt~234402: Discovery of Luminous X-ray Ignition}
   \subtitle{}
   \author{
        D. Homan\inst{\ref{aip}}
        \and M. Krumpe\inst{\ref{aip}}
        \and A. Markowitz\inst{\ref{ncap},\ref{ucsd}}
        \and T. Saha\inst{\ref{ncap}}
        \and A. Gokus\inst{\ref{wustl},\ref{remeis},\ref{wuerzburg}}
        \and E. Partington\inst{\ref{wayne}}
        \and G. Lamer\inst{\ref{aip}}
        \and A. Malyali\inst{\ref{mpe}}
        \and Z. Liu\inst{\ref{mpe}}
        \and A. Rau\inst{\ref{mpe}}
        \and I. Grotova\inst{\ref{mpe}}
         \and E. M. Cackett\inst{\ref{wayne}}
        \and D.A.H. Buckley\inst{\ref{saao},\ref{uct},\ref{ufs}}
        \and S. Ciroi\inst{\ref{asiago}}
        \and F. Di Mille\inst{\ref{campanas}}
        \and K. Gendreau\inst{\ref{nasa}}
        \and M. Gromadzki\inst{\ref{aouw}}
        \and S. Krishnan\inst{\ref{ncap}}
        \and M. Schramm\inst{\ref{naoj}}
        \and J.F. Steiner\inst{\ref{cfa}}
    }
   \institute{
        Leibniz-Institut für Astrophysik Potsdam, An der Sternwarte 16, 14482 Potsdam, Germany\label{aip}
        \and Nicolaus Copernicus Astronomical Center, Polish Academy of Sciences, ul.\ Bartycka 18, 00-716 Warszawa, Poland\label{ncap}
        \and University of California, San Diego, Center for Astrophysics and Space Sciences, MC 0424, La Jolla, CA, 92093-0424, USA\label{ucsd}
        \and Department of Physics, Washington University in St. Louis, One Brookings Drive, St. Louis, MO 63130, USA\label{wustl}
        \and Dr. Karl Remeis-Observatory \& ECAP, Friedrich-Alexander-Universit\"at Erlangen-N\"urnberg, Sternwartstr. 7, 96049 Bamberg, Germany\label{remeis}
        \and Lehrstuhl f\"ur Astronomie, Universit\"at W\"urzburg, Emil-Fischer-Straße 31, 97074 W\"urzburg, Germany\label{wuerzburg}
        \and Wayne State University, Department of Physics \& Astronomy, 666 W. Hancock St, Detroit, MI 48201, USA\label{wayne}
        \and Max-Planck-Institut für extraterrestrische Physik, Gießenbachstraße, 85748 Garching, Germany\label{mpe}
        \and South African Astronomical Observatory, PO Box 9, Observatory, Cape Town 7935, South Africa\label{saao}
        \and Department of Astronomy, University of Cape Town, Private Bag X3, Rondebosch 7701, South Africa\label{uct}
        \and Department of Physics, University of the Free State, PO Box 339, Bloemfontein 9300, South Africa\label{ufs}
        \and Department of Physics and Astronomy, University of Padova, Via F. Marzolo 8, I-35131 Padova, Italy\label{asiago}
        \and Las Campanas Observatory – Carnegie Institution for Science, Colina el Pino, Casilla 601, La Serena, Chile\label{campanas}
        \and Astrophysics Science Division, NASA Goddard Space Flight Center, Greenbelt, MD 20771, USA\label{nasa}
        \and Astronomical Observatory, University of Warsaw, Al. Ujazdowskie 4, PL-00-478 Warsaw, Poland\label{aouw}
        \and Graduate School of Science and Engineering, Saitama University, 255 Shimo-Okubo, Sakura-ku, Saitama City, Saitama 338-8570, Japan\label{naoj}
        \and Center for Astrophysics, Harvard \& Smithsonian, 60 Garden St, Cambridge, MA 02138, U.S.A.\label{cfa} 
    }
   \date{\today}

\abstract{In November 2020, a new, bright object, eRASSt~J234402.9$-$352640, was discovered in the second all-sky survey of \textit{SRG}/eROSITA. The object brightened by a factor of at least 150 in 0.2--2.0 keV flux compared to an upper limit found six months previous, reaching an observed peak of $1.76_{-0.24}^{+0.03} \times 10^{-11}$ erg~cm$^{-2}$~s$^{-1}$. The X-ray ignition is associated with a galaxy at $z=0.10$, making the peak luminosity log$_{10}(L_{\rm 0.2-2keV}/[\textrm{erg s}^{-1}])$=$44.7\pm0.1$. Around the time of the rise in X-ray flux, the nucleus of the galaxy brightened by approximately 3 mag. in optical photometry, after correcting for the host contribution. We present X-ray follow-up data from \textit{Swift}, \textit{XMM-Newton}, and \textit{NICER}, which reveal a very soft spectrum as well as strong 0.2--2.0~keV flux variability on multiple timescales. Optical spectra taken in the weeks after the ignition event show a blue continuum with broad, asymmetric Balmer emission lines, and high-ionisation ([OIII]$\lambda\lambda$4959,5007) and low-ionisation ([NII]$\lambda$6585, [SII]$\lambda\lambda$6716,6731) narrow emission lines. Following the peak in the optical light curve, the X-ray, UV, and optical photometry all show a rapid decline. The X-ray light curve shows a decrease in luminosity of $\sim$0.45 over 33 days and the UV shows a drop of $\sim$0.35 over the same period. eRASSt~J234402.9$-$352640 also shows a brightening in the mid-infrared, likely powered by a dust echo of the luminous ignition. We find no evidence in \textit{Fermi}-LAT $\gamma$-ray data for jet-like emission. The event displays characteristics of a tidal disruption event (TDE) as well as of an active galactic nucleus (AGN), complicating the classification of this transient. Based on the softness of the X-ray spectrum, the presence of high-ionisation optical emission lines, and the likely infrared echo, we find that a TDE within a turned-off AGN best matches our observations.}

   \keywords{X-rays: individuals: eRASSt~J234402.9$-$352640 --- Accretion, accretion disks --- Galaxies: active}

   \maketitle
%

\section{Introduction}\label{sec:intro}
All massive galaxies are thought to harbour a supermassive black hole (SMBH) at their centre \citep[][]{SOLTAN_1982}. A SMBH can power strong emission through accretion of surrounding matter, both on long timescales and in shorter-lived outbursts. Active galactic nuclei (AGN) represent the more continuous form of accretion and have a strong impact on the evolution of their host galaxies through various feedback mechanisms \citep[e.g.][]{KORMENDY_2013,HECKMAN_2014}. Variability is inherent to AGN emission, across the electromagnetic spectrum and on all timescales. X-ray observations show that between 1\% and 4\% of local galaxies host an AGN \citep[e.g.][]{HAGGARD_2010,BIRCHALL_2022}. It is likely that active accretion is intermittent over a galaxy's lifetime; however, AGN duty cycles lack strong observational constraints \citep[][]{NOVAK_2011,SCHAWINKSI_2015,SHEN_2021}. Stochastic (i.e. aperiodic) optical continuum variability in AGN has been found to be of the order of 10--20\% of the flux on a timescale of months \citep[e.g.][]{KELLY_2009,MACLEOD_2013}, increasing to 2--3 times the flux on a timescale of years \citep[e.g.][]{UTTLEY_2003,BREEDT_2009,BREEDT_2010}. More extreme variability has also been observed, including large outbursts in both X-ray and optical emission \citep[e.g.][]{BRANDT_1995,SHAPPEE_2014,TRAKHTENBROT_2019B}. In recent years, studies of large samples of AGN have revealed that optical variability up to 1 mag is not uncommon and can be found in up to 50\% of AGN on a timescale of a decade \citep[][]{LAWRENCE_2016,RUMBAUGH_2018,GRAHAM_2020}. Whether strong variability is the extreme end of `regular' AGN variability, or is representative of a different type of process, such as a change in accretion mode, is still an open question \citep[e.g.][]{RUAN_2019}. 

An inherently more short-term SMBH accretion event is caused by the tidal disruption of a star followed by the subsequent fallback and accretion of the debris, referred to as a tidal disruption event \citep[TDE;][]{REES_1988,PHINNEY_1989}. These events were initially predicted to be observed as large-amplitude, ultra-soft X-ray flares originating from galactic nuclei \citep{REES_1988}. The first such candidates were observed in the R\"{o}ntgensatellit \citep[\textit{ROSAT};][]{TRUEMPER_1992} All-Sky survey \citep[e.g.][]{BADE_1996,GRUPE_1999,KOMOSSA_1999,KOMOSSA_199909,GREINER_2000,DONLEY_2002}. In recent years, the majority of TDE candidates have been found through optical surveys \citep[e.g.][]{VANVELZEN_2011,GEZARI_2017,HUNG_2017,VANVELZEN_2021}. Optically selected TDEs are characterised by light curves that show a rapid rise in brightness followed by a relatively smooth decay. Follow-up optical spectroscopy of TDE candidates generally reveals transient, blue continua (blackbody temperatures $\sim$10$^4$ K), and in some cases broad emission lines \citep[e.g. Balmer, He~II, and/or Bowen lines;][and references therein]{CHARALAMPOPOULOS_2022}. 
As the number of TDE candidates has grown, it has become increasingly clear that these systems show a broad range of observed behaviours, such as the presence or absence of optical emission lines \citep[][]{LELOUDAS_2019,VANVELZEN_2021}, the presence or absence of radio emission \citep[][]{ALEXANDER_2020}, the presence or absence of X-ray emission \citep[][]{HOLOIEN_2016,AUCHETTL_2018}, as well as differing rates of decay in the post-peak phase \citep[][]{VANVELZEN_2021}, and a candidate with a double-peaked optical light curve \citep[][]{MALYALI_2021,CHEN_2022}.

Due to the broad phenomenological range in both AGN variability and TDEs, distinguishing these different classes of events has proven difficult in many cases \citep[][]{MERLONI_2015,AUCHETTL_2018}. Fundamentally, they both represent a rapid change in the accretion rate, `powering up' the central engine, although the precise accretion processes could differ among TDEs \citep[for a recent review, see][]{GEZARI_2021}. In recent years, a number of studies have been published concerning strong, `flaring' ignitions in AGN, associating them with either extreme AGN variability \citep[e.g.][]{GEZARI_2017,YAN_2019,TRAKHTENBROT_2019A,FREDERICK_2021} or with a TDE \citep[e.g.][]{VANVELZEN_2016,BLANCHARD_2017,LIU_2020,BRIGHTMAN_2021}. 
A combination of AGN-like and TDE-like features in other objects has led to their identification as transient events within AGN \citep[][]{RICCI_2020,HOLOIEN_2021}, or has made a definitive classification impossible without additional data \citep[e.g.][]{NEUSTADT_2020,MALYALI_2021,HINKLE_2022}.

The Extended ROentgen Survey with an Imaging Telescope Array \cite[eROSITA;][]{PREDEHL_2021}, aboard the \textit{Spectrum Roentgen/Gamma} (\textit{SRG}) spacecraft \citep{SUNYAEV_2021}, has greatly expanded the search for transients in the X-ray sky. A number of new TDE candidates have already been detected by \textit{SRG}/eROSITA \citep[][]{SAZONOV_2021,MALYALI_2021}. In this paper, we report on the initial discovery and multi-wavelength follow-up of eRASSt~J234402.9$-$352640 (hereafter J234402), which was detected in the second eROSITA all-sky survey.  A significant increase in the flux from the same source measured by \textit{Gaia} on 2020-10-14 was reported by \citet[][]{GAIA_ALERT_2020} in TNS Astronomical Transient Report (\#85552). In this paper we present our initial findings and conclusions, as we pursue ongoing follow-up of this object. The data and results presented here pertain to the period up to February 2021, when further observations of the object had to be paused due to Sun block. We will present the results of the resumed, post-Sun-block observations in follow-up work (Malyali et al., in prep.). 

The paper is structured as follows. In Section~\ref{sec:xray} we present our X-ray observations. In Section~\ref{sec:mw} we describe our datasets in other wavebands: Section~\ref{sec:mw_phot} covers the available UV, optical, and IR photometry; Section~\ref{sec:mw_spec} provides an overview of our optical spectroscopic follow-up; in Section~\ref{sec:mw_fermi} we describe our analysis of \textit{Fermi}-LAT $\gamma$-ray data. We discuss our results in Section~\ref{sec:discussion}, within the context of extreme AGN variability and a TDE. We summarise our conclusions in Section~\ref{sec:conclusion}. Throughout this paper, we assume a flat cosmology with $H_0 = 70$\,km s$^{-1}$ Mpc$^{-1}$, $\Omega_M = 0.3$, and $\Omega_{\Lambda} = 0.7$ \citep[][]{HINSHAW_2013}.  In this cosmology, 1\arcsec\ corresponds to 1.8 kpc at the redshift of eRASSt~J234402$-$352640 ($z=0.10$; light travel time 1.3 Gyr), where the associated luminosity distance is 460 \,Mpc. All uncertainties represent a 1$\sigma$ (68.3\% for a single parameter) confidence interval unless stated otherwise.


\section{X-ray observations \& analysis}\label{sec:xray}
\begin{figure*}
    \centering
    \includegraphics[width=.8\linewidth]{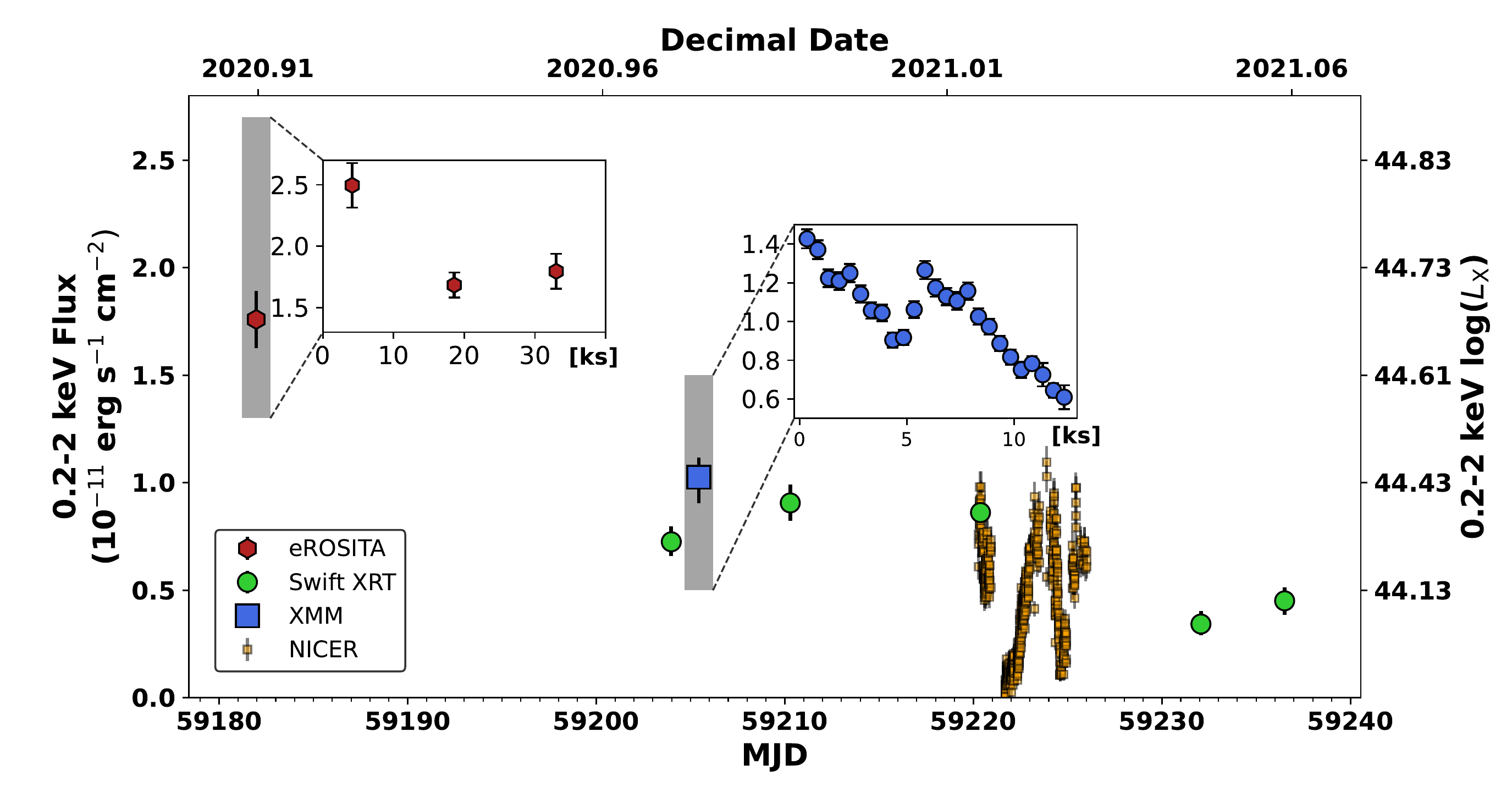}
    \caption{Overview of X-ray data collected on J234402. The light curve starts from the detection of the ignition event with eROSITA in November 2020. The 3$\sigma$ upper limit of the eRASS1 scan in May 2020 was 1.1 $\times$ 10$^{-13}$ erg~cm$^{-2}$~s$^{-1}$ (log$_{10}(L_\textrm{X}/[\rm {erg\,s^{-1}}])=41.8$). The flux evolution over time is significant, showing both an overall decrease and large fluctuations on shorter timescales. To illustrate the magnitude of the variability on the shortest timescales, the light curves of individual eRASS2 scans and for the \textit{XMM-Newton} EPIC-pn data, binned to 500s, are included in insets. The vertical extent of the grey boxes around the eROSITA and \textit{XMM-Newton} markers shows the size of the y-axes of the insets within the larger plot. 
    In the inset of the eROSITA light curve the horizontal axis marks the time in ks since MJD 59181.75, and in the inset of the \textit{XMM-Newton} light curve it marks the time in ks since MJD 59205.44. On the right vertical axis we show the log$_{10}(L_\textrm{X}/[\rm {erg\,s^{-1}}])$ values associated with the observed flux. To calculate these values, we made use of the best-fit model (2BB, Section~\ref{sec:xray_xmm}).}
    \label{fig:xray_combined}
\end{figure*}

Following detection by eROSITA, X-ray-follow-up observations of J234402 took place between November 2020 and January 2021, after which the object was positioned too close to the Sun for follow-up. The follow-up comprised a \textit{Swift} monitoring programme, an \textit{XMM-Newton} pointing, and an extended \textit{NICER} observation covering 18 days. The combined dataset is presented in Figure~\ref{fig:xray_combined}. For the analysis in this paper, we use the following parameters and measurements unless otherwise stated: a Galactic total ({H}\textsc{I} + H$_2$) hydrogen column density N$_\mathrm{H} = 1.2\times10^{20}$~cm$^{-2}$ \citep{WILLINGALE_2013}, the cosmic abundances of \citet{WILMS_2000}, and the photoelectric absorption cross sections provided by \citet{VERNER_1996}.

We checked the Second ROSAT Source Catalogue \citep[2RXS;][]{BOLLER_2016} for a counterpart that could have been found between June 1990 and July 1991 during the ROSAT All-Sky Survey. No point source is detected within a 15 arcminute radius of J234402. The average flux upper limit for inclusion in 2RXS is estimated by Boller et al. to be approximately $10^{-13}$~erg~cm$^{-2}$~s$^{-1}$ in the 0.1--2.0~keV band. For a redshift of 0.10 and an assumed photon index of 2.0, we find an upper limit of $L_\textrm{0.1--2 keV}\sim2\times10^{42}$ erg~s$^{-1}$ for the ROSAT epoch. The limit on the absorbed luminosity indicates there was no strong X-ray emission --~which would have been indicative of an AGN~--  at the time of the ROSAT survey. No serendipitous archival observations of the target were found for \textit{XMM-Newton}, \textit{Swift}, or \textit{Chandra}.

\begin{table*}
\newcolumntype{a}{>{\columncolor{red}}c}
\renewcommand*{\arraystretch}{1.6}
    \centering
    \caption{
    Overview of the X-ray fluxes measured for J234402.
    }
    \begin{tabular}{p{2cm}ccc|cc|cc}\toprule
         MJD & GTI$^a$ & Observatory & $f_{0.2-2\,\rm keV}^b$ & $\Gamma^c$ & TS/\textit{dof} & $k_\textrm{B}T^d$ (eV) & TS/\textit{dof}$^e$ \\
         \midrule
         58993.76 & 0.2 & eROSITA (eRASS1) & $<0.11$ & -- & -- & --  & --\\
         59181.97 & 0.1 & eROSITA (eRASS2) & $17.59^{+0.28}_{-2.36}$ & $4.7\pm0.1$ & 110.5/110 & $72\pm1.4$ & 184.7/110\\
         59203.98 & 0.9 & \textit{Swift}-XRT & $7.25^{+0.72}_{-0.66}$ & $4.0\pm0.1$ & 34.1/47 & $93^{+17}_{-15}$ & 32.7/47\\
         59205.44 & 8.1 & \textit{XMM-Newton} & $10.26^{+0.09}_{-0.12}$& 5.3* & 1036.5/421 & 74* & 1339.2/421 \\
         59210.29 & 1.1 & \textit{Swift}-XRT & $9.06^{+0.86}_{-0.83}$ & $3.9\pm0.7$ & 38.7/47 & $94^{+14}_{-12}$ & 40.8/47\\
         59220.38 & 9.9 & \textit{Swift}-XRT & $8.61^{+1.16}_{-1.10}$ & $4.4^{+1.1}_{-0.9}$ & 40.4/47 & $79^{+20}_{-18}$ & 41.4/47\\
         59220.28 -- \newline 59226.02 & \multirow{1.7}{*}{18.5} & \multirow{1.7}{*}{NICER} & \multirow{1.7}{*}{$4.39^{+0.03}_{-0.05}$} & \multirow{1.7}{*}{$5.1^{+0.0}_{-0.0}$} & \multirow{1.7}{*}{1643.5/136} & \multirow{1.7}{*}{$73^{+0.3}_{-0.3}$} & \multirow{1.7}{*}{644.6/136}\\
         59232.07 & 0.8 & \textit{Swift}-XRT & $3.43^{+0.62}_{-0.52}$ & $4.5*$ & 41.8/47 &  $82*$ & 48.5/47\\
         59236.51 & 1.0 & \textit{Swift}-XRT & $4.51^{+0.63}_{-0.65}$ & $5.6^{+1.9}_{-1.4}$ & 48.2/47 & $63^{+18}_{-17}$ & 45.1/47\\
     \bottomrule
    \end{tabular}
    \flushleft{\scriptsize{
    a) Good time interval in kiloseconds, for eRASS1 we list the total exposure time.\\
    b) Fluxes are given in units of $10^{-12}$ erg cm$^{-2}$ s$^{-1}$ and represent the integrated flux over the range 0.2 to 2 keV. The fluxes are derived from the best-fitting models for each dataset.\\
    c) Slope of the best-fit single power-law model. The errors on the fit parameters are somewhat unreliable, as this simple model proves to be a poor fit to the spectrum for J234402. For the \textit{XMM-Newton} observation and the \textit{Swift-XRT} observation on 18-01-2021 the fits are too poor to estimate an uncertainty, these measurements have been marked with an *.\\
    d) Temperature of the best-fit single blackbody model.\\
    e) Test statistic (TS) compared to the number of degrees of freedom. The test statistic is $\chi^2$ for the \textit{XMM-Newton} observation, and Cash for eROSITA and \textit{Swift}-XRT data.\\
    }}
    \label{tab:xray_lc}
\end{table*}

\subsection{SRG/eROSITA}\label{sec:xray_ero}
The X-ray ignition in J234402 was detected by eROSITA through a comparison of sources detected in the first and second eROSITA All-Sky Survey (eRASS1 and eRASS2). The eROSITA position of the event is R.A., Dec. = 23:44:02.9, $-$35:26:39.4 (J2000) with a total uncertainty of 1.0 arcseconds, where we have included a 0.6 arcseconds systematic uncertainty. We identified a point source at a distance of 2.2 arcseconds with coordinates R.A., Dec. = 23:44:02.9, $-$35:26:41.6 in the Gaia eDR3 catalogue \citep[][]{GAIA_EDR3} as the optical counterpart. The eROSITA data presented in this paper were reduced using eSASSusers\_211214 \citep[][]{BRUNNER_2022} and we made use of HEASOFT v6.28 (XSPEC 12.11.1). 

\begin{table}[ht]
    \centering
    \caption{
    Fluxes for individual scans of eRASS2 epoch.
    }
    \begin{tabular}{lcc}
        \toprule
        MJD & Hours$^a$ & $f_{0.2-2\,\rm keV}^b$ \\
        \midrule
        59181.798 & 0 & 25.0$\pm$1.8 \\
        59181.965 & 4.0 & 16.8$\pm$1.0 \\
        59182.132 & 8.0 & 18.0$\pm$1.4 \\
        \bottomrule
    \end{tabular}
    \flushleft{\scriptsize{
    a) Elapsed time since first scan of eRASS2.\\
    b) Fluxes and their 1$\sigma$ uncertainties are given in units of $10^{-12}$ erg cm$^{-2}$ s$^{-1}$.\\
    }}
    \label{tab:erass2fluxes}
\end{table}

The object was not detected in eRASS1 (24--25 May 2020), with a 3$\sigma$ sensitivity limit of $f_{3\sigma,0.2-2{\rm keV}}= 1.1 \times 10^{-13}$ erg~cm$^{-2}$~s$^{-1}$. However, in eRASS2 (27--29 November 2020) the source was discovered with a 0.2--2 keV flux of $1.76_{-0.24}^{+0.03} \times 10^{-11}$ erg cm$^{-2}$ s$^{-1}$. The eROSITA fluxes are listed in Table~\ref{tab:xray_lc}, which provides an overview of all X-ray fluxes measured in our follow-up programme. A single eRASS epoch comprises multiple scans that each last approximately 40 seconds, covering the same sky position at intervals of 4 hours. As the satellite was affected by a technical issue, there are only three individual scans for which we could extract flux data within the eRASS2 epoch. The values for the individual scans are shown in Table~\ref{tab:erass2fluxes}. The eRASS2 fluxes show remarkable variability, dropping by $\sim$1/3 over 4 hours between the first and second scan, with a 4$\sigma$ significance for this change. 

\begin{figure}
    \centering
    \includegraphics[width=\linewidth]{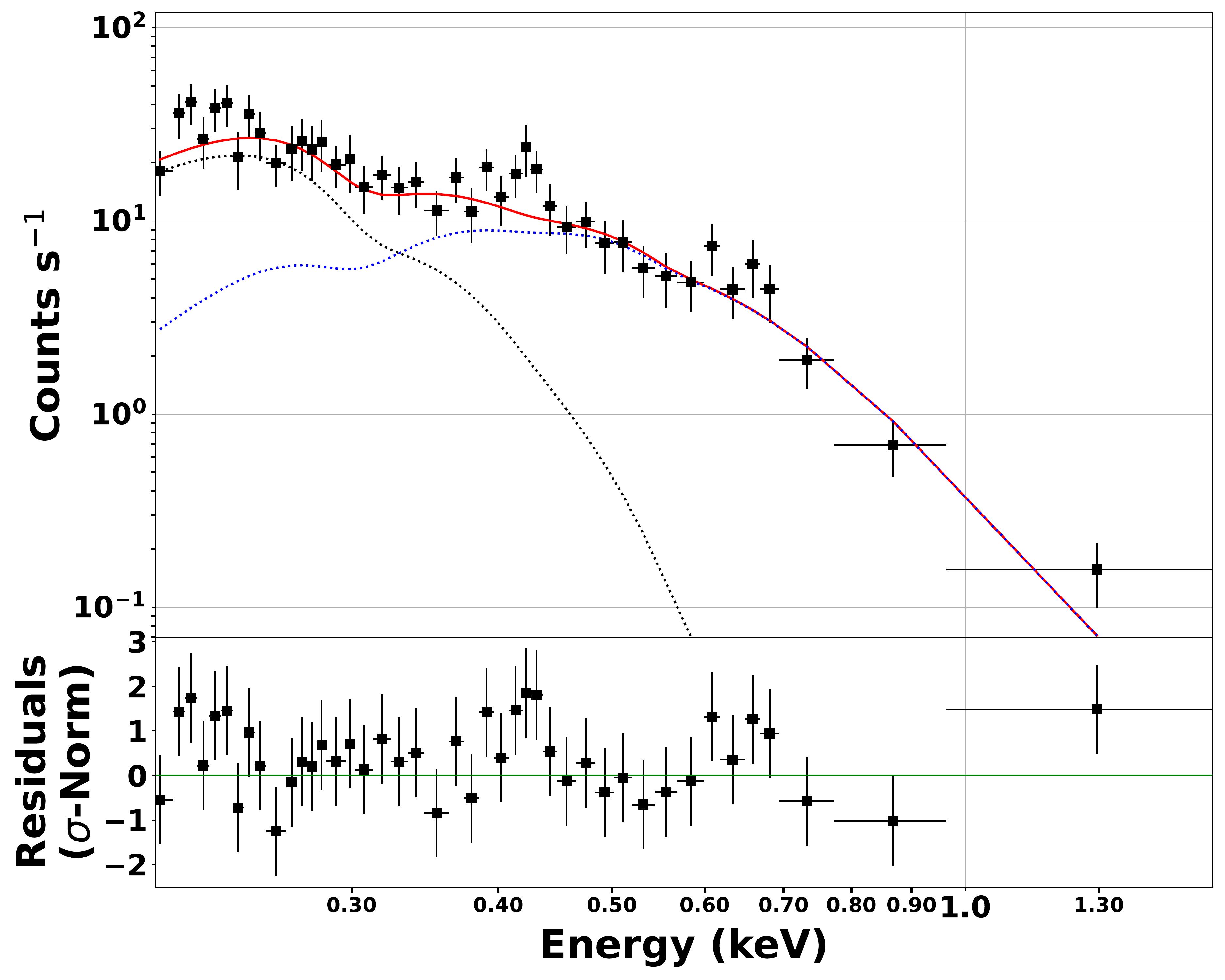}
    \caption{Binned eROSITA spectrum of J234402. The spectrum for the eRASS2 observation (28--29 November 2020) is very soft, a characteristic confirmed in all our X-ray observations. Included in red (solid line) is the best-fit model, consisting of two redshifted black bodies (dotted lines), corrected for Galactic absorption.}
    \label{fig:ero_spec}
\end{figure}

The combined spectrum for the eRASS2 observation is shown in Figure~\ref{fig:ero_spec}. The spectrum is soft, with most counts below 0.5 keV. We fitted the eRASS2 spectrum with a range of models and find that a model consisting of two blackbodies modified by Galactic absorption (\textsc{tbabs*(zbbody+zbbody)}) provides the best fit, with $\chi^2$/\textit{dof} = 140.7/108. The blackbody temperatures in this fit are $100\pm 5$~eV and $38\pm 4$~eV. The best-fit model is included in Figure~\ref{fig:ero_spec}. 

\subsection{XMM-Newton}\label{sec:xray_xmm}
The \textit{XMM-Newton} observation of J234402 took place on 22 December 2020 (MJD 59203), with an exposure of 14 ks. The data were reduced with the SAS package (18.0.0) and HEASOFT (v6.24) using the standard settings for point sources. Small window mode was used and pile-up does not affect the observations. We created a standard source and background spectrum for each of pn, MOS1, and MOS2. 
We found total good time intervals of 8.1 ks ($\sim$34,000 pn net counts). The spectra were grouped with a minimum binning of 20 counts per bin. For the pn, we extracted single (pattern 0) and double (patterns 1--4) spectra separately (hereafter pn0 and pn14), as the double-events spectrum has calibration issues below 0.4 keV due to the use of small window mode. 

No significant detection above 1.2~keV was made in any camera. We therefore extracted the spectra in the range 0.2--1.2 keV for pn0, MOS1, and MOS2, and in 0.4--1.2 keV for pn14. The counts show significant variability, dropping by a factor of two over $\sim$3 hours. We quantify the variability using the fractional rms variability amplitude \citep[][]{VAUGHAN_2003}, and find $F_{\rm var} = 20.2\pm 0.9\%$. Although the X-ray flux changed significantly over the course of the \textit{XMM-Newton} exposure, we do not detect any significant changes in the X-ray spectrum over this same period. 

We tested a range of X-ray spectral models using the nested sampling package \textsc{multinest} v.3.10 \citep{SKILLING_2004,FEROZ_2009} via the Bayesian X-ray Analysis and PyMultinest packages for \textsc{Xspec} \citep[BXA, version 3.31, using Xspec version 12.10.1;][]{BUCHNER_2014}. We used default arguments for \textsc{multinest} (400 live points and a sampling efficiency of 0.8), and assumed uniform or log-normal initial prior distributions. We used the 5th and 95th percentiles of the posterior distribution for Bayesian fits. We report Bayesian evidence $Z$, to compare goodness of fit between models\footnote{$Z$ is defined following \citet{BUCHNER_2014}, $\S5.2$.}. 

Multiple one-component models yield poor fits, but reveal both the extreme softness of the spectrum and its strong convex spectral curvature. Physically motivated models based on disk emission, such as \textsc{diskpbb} and \textsc{comptt} achieve reasonable results, provided the seed photon temperature and (for \textsc{comptt}) the electron temperature is of the order of 50--100 eV. The overall best-fitting model is the purely phenomenological log-parabola (\textsc{zlogpar}). The best-fitting physically motivated model is a double blackbody (\textsc{zbbody+zbbody}, 2BB), with $k_{\rm B}T_1 = 53 \pm 2$ eV and $k_{\rm B}T_2 = 109^{+4}_{-3}$ eV. We also tested for the presence of ionised absorption, by making use of the \textsc{warmabs} model. Combinations \textsc{warmabs*zbbody} and \textsc{warmabs*powerlaw} resulted in poor fits, as the absorption could not account for the spectral curvature. Combinations of \textsc{warmabs} with well-fitting models such as \textsc{zlogpar} resulted in a negligible improvement in goodness of fit. More complex models including Comptonised components also yield acceptable fits, with temperatures in the same order of magnitude as 2BB. An overview of our fitting results is shown in the Appendix Table~\ref{tab:xray_xmmfit}.

\begin{figure}
\includegraphics[width=0.9\columnwidth]{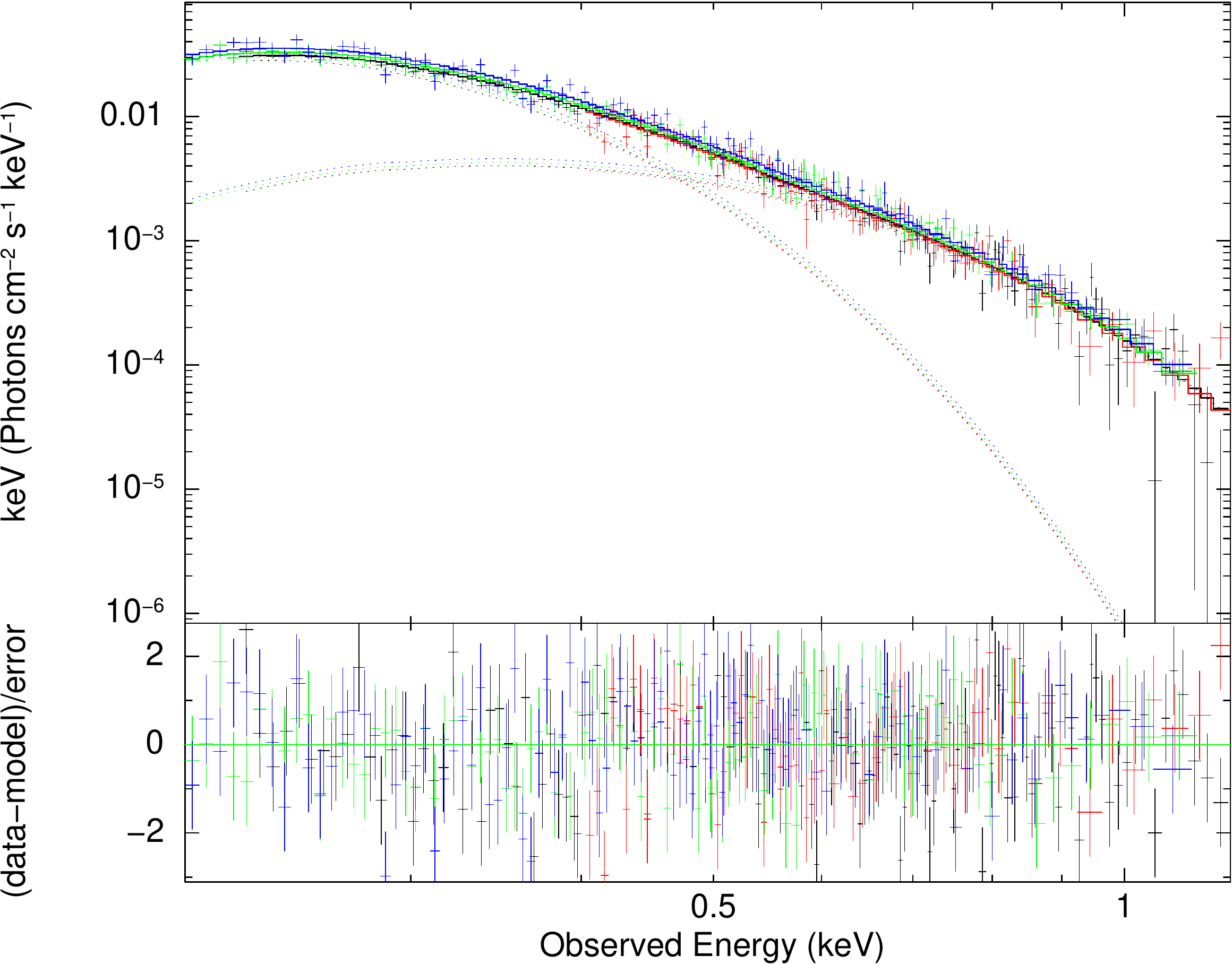}
\caption{Unfolded model spectrum from \textit{XMM-Newton}. The spectrum was unfolded using the best-fit dual blackbody model (2BB), with blackbody temperatures of $53 \pm 2$~eV and $109^{+4}_{-3}$~eV. The two blackbody components blend into one another, matching the soft and featureless X-ray spectrum. Black, red, green, and blue points denote pn0, pn14, MOS1, and MOS2 data, respectively.}
\label{fig:xmmeufdel}
\end{figure}

Based on our fitting results, it is not clear if one- or two-component models are generally preferred. However, the spectral data seem to lack an obvious `inflection' point indicating distinct emission components, and all of the two-component fits have components blending slowly into each other. Of the models based on physical parameters, we find that the 2BB model provides the best fit in relation to the number of free parameters. We therefore adopt 2BB as our optimal model. An unfolded model spectrum for our dual-blackbody model is plotted in Figure~\ref{fig:xmmeufdel}. The 0.2--2.0~keV model flux for the full observation is $1.026^{+0.009}_{-0.012}\times10^{-11}$~erg~cm$^{-2}$~s$^{-1}$ in our best-fit model (1LOGPAR). Using 2BB we find a very similar value.

\subsection{\textit{Swift}-XRT}
\label{sec:xray_swift}
The \textit{Neil Gehrels Swift Observatory} \citep[][]{GEHRELS_2004} observed J234402 on five occasions (PI A. Malyali): 20 and 27 December 2020, followed by 6, 18, and 22 January 2021 (MJDs 59203, 59210, 59220, 59232, and 59236). Each exposure lasted approximately 1~ks. We present spectral data from the X-ray Telescope (XRT) and photometry data from the Ultraviolet/Optical Telescope (UVOT; the UV data are described in Section~\ref{sec:mw_opt_swift}). All XRT observations used Photon Counting mode. Event files were calibrated using \textsc{xrtpipeline} in HEASOFT v.6.28 and the latest calibration files. Source spectra were extracted from a circular region of radius 20 pixels (47$\arcsec$). Backgrounds were extracted from annular regions with inner and outer radii 30 and 36 pixels (70$\arcsec$ and 84$\arcsec$), which were selected to be free from background sources. We used the latest response matrix in the calibration database. Ancillary response files were generated using \textsc{xrtmkarf}.

We fitted the \textit{Swift}-XRT spectra in the range 0.2--2.0 keV using \textsc{Xspec}, making use of Cash statistics, due to the relatively low photon count. The spectra were re-binned to a minimum of 15 counts per bin. We fitted a single power-law, a single blackbody, and a dual blackbody (2BB). For 2BB we kept the temperatures and the ratio of the normalisations frozen at the best-fit values of the \textit{XMM-Newton} EPIC-pn fit. Untying the normalisations yielded a negligible improvement in goodness of fit. For the single-power-law and single-blackbody fits to the observation of MJD 59232 the low number of counts required us to freeze the values $\Gamma$ and $k_{\rm B}T$, respectively, to the average values from the fits to the other four XRT spectra. The derived fluxes are included in Table~\ref{tab:xray_lc}, and a full overview of the results of the model fits is presented in the Appendix Table~\ref{tab:xrtfits}. The XRT data do not allow us to significantly prefer one model over another. For consistency, we derived the 0.2--2.0 keV \textit{Swift}-XRT fluxes using the best-fit 2BB model.

\subsection{NICER}
\label{sec:xray_nicer}
J234402 was observed with \textit{NICER} in the period covering 5 January 2021 to 22 January 2021 (MJD 58853--58870). Due to high levels of optical loading, the final good time intervals where the background could be reliably subtracted cover the period 6 to 11 January (MJD 58854--58859, see Appendix~\ref{sec:app_xray_nicer} for further discussion). We made use of the \textit{NICER} reduction pipeline to produce the standard RMF, ARF, and background files. We fitted the \textit{NICER} data using \textsc{Xspec} v12.12.0, which is part of HEASOFT v6.28. The data have been re-binned to a minimum of 25 counts per bin and we use $\chi^2$-statistics. The spectra were fit in the 0.3--2 keV energy range.

\subsubsection{\textit{NICER} light curve}
\begin{figure}
    \centering
    \includegraphics[width=\linewidth]{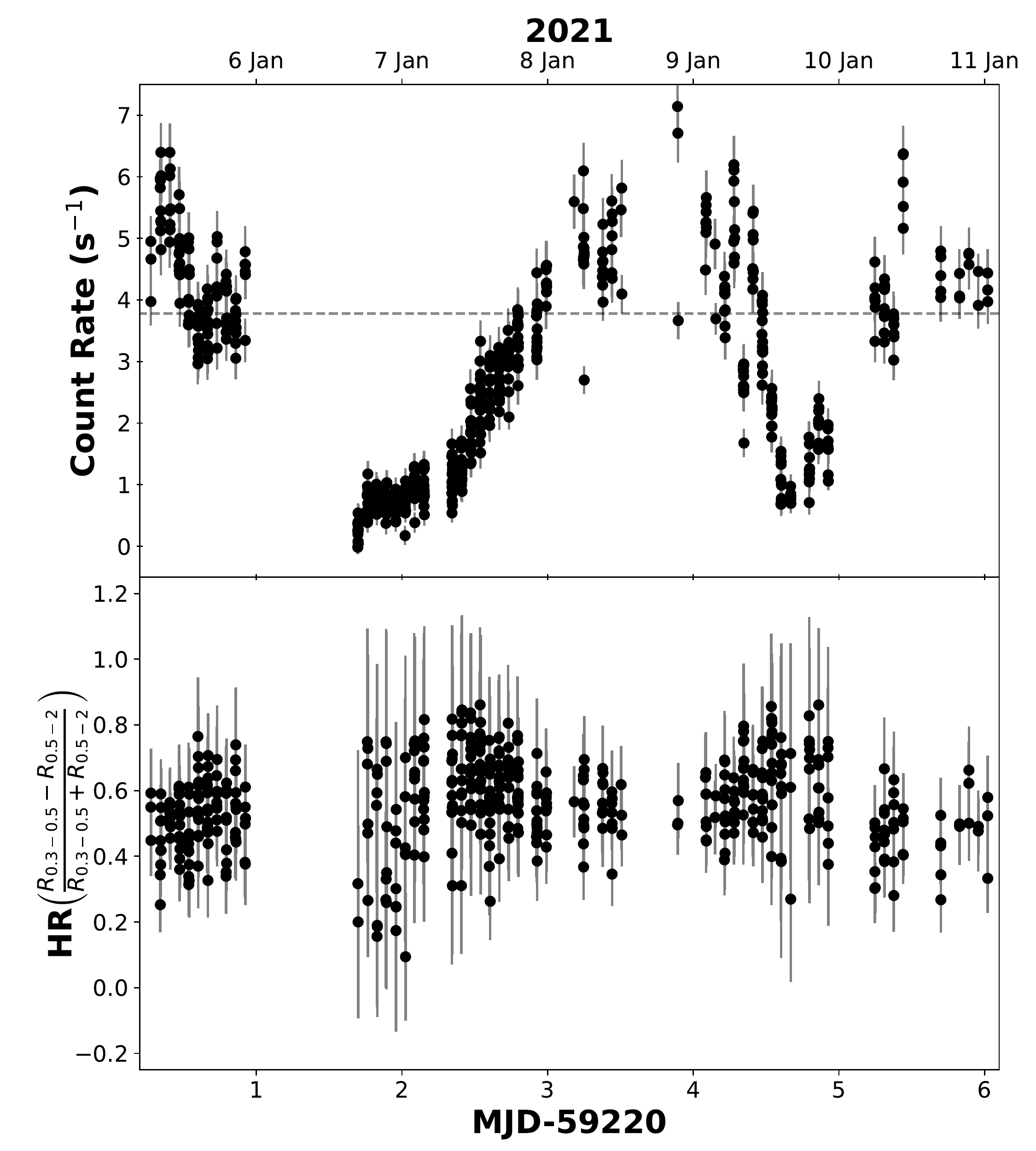}
    \caption{\textit{NICER} light curve and hardness ratio of J234402. \textit{top}: 0.3--2 keV \textit{NICER} light curve. The light curve shows significant variability on a timescale of days. The sharp turns up and down are consistent with the rapid variability observed in the \textit{XMM-Newton} and eROSITA data. The dashed line indicates the cut-off between the low-state and high-state intervals at 3.78 counts~s$^{-1}$, as discussed in Section~\ref{sec:xray_nicer_spec}. \textit{bottom}: hardness ratio (HR) in \textit{NICER} light curve. We define the HR using the 0.3--0.5 keV and the 0.5--2 keV count rates ($R_{0.3-0.5}$ and $R_{0.5-2}$, respectively), as defined in the text. We detect no significant change over time.}
    \label{fig:nicer_lc}
\end{figure}

The counts light curve for the energy band 0.3--2~keV is presented in Figure~\ref{fig:nicer_lc}. The source clearly shows strong variability on a timescale of days. We find $F_{\rm var}=55.4\pm0.5\% $. Although the light curve shows a pattern of repeated ups and downs, we would caution against an interpretation as a (quasi-)periodic oscillation based on these data alone. There is extensive literature on the appearance of temporary sinusoid-like signals in data generated by stochastic red noise processes, as well as mistaken claims of detection of periodic signals \citep[for a discussion, see e.g.][]{PRESS_1978,VAUGHAN_2016}. We note that we do not exclude the possibility of a periodic signal in the \textit{NICER} data, but stress that the current sampling and duration do not allow for a claim of detection, especially given that such a claim would be based on observing a very small number of putative cycles. 

\subsubsection{\textit{NICER} spectra}\label{sec:xray_nicer_spec}
\begin{figure}
    \centering
    \includegraphics[width=\linewidth]{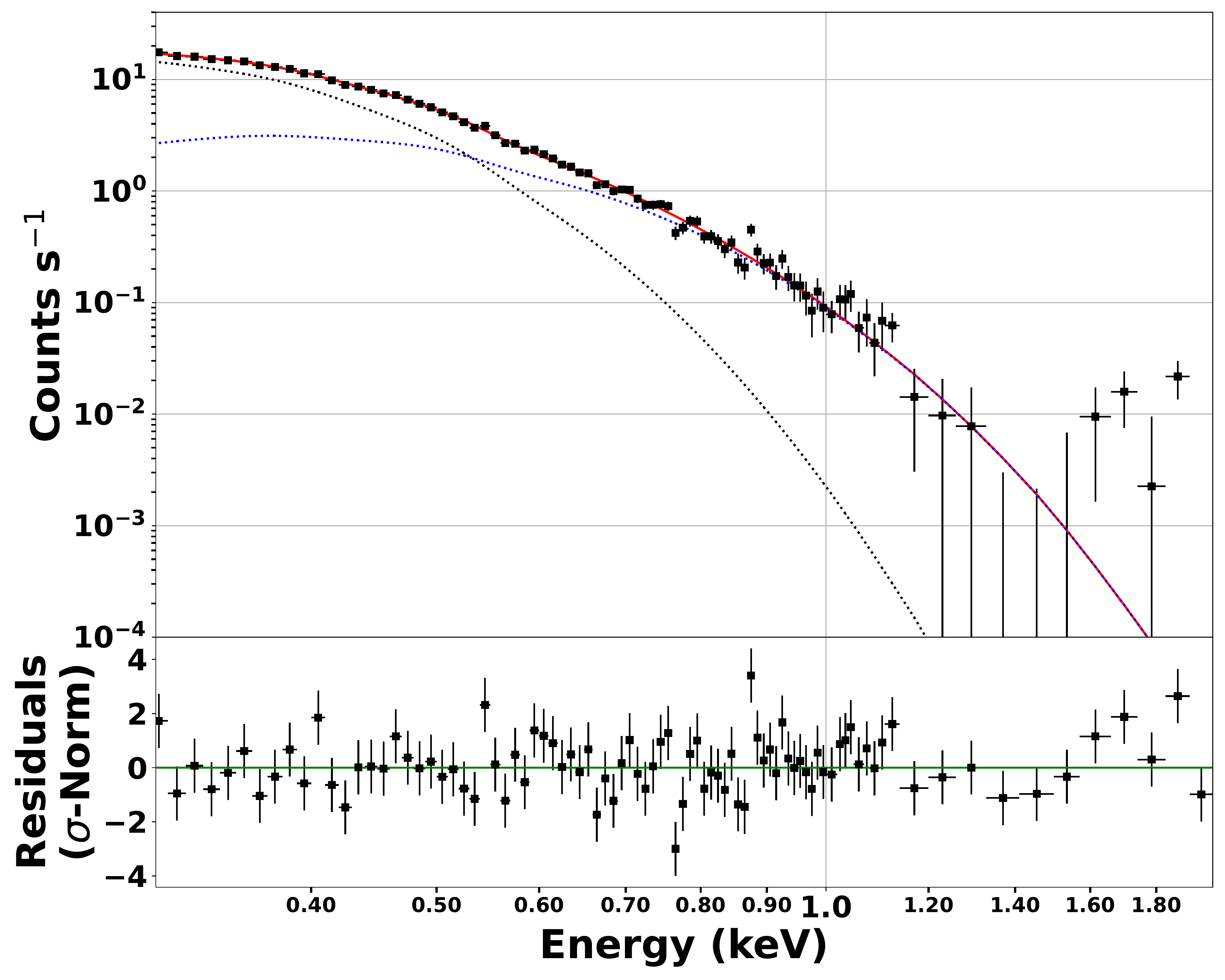}
    \caption{Spectrum from \textit{NICER} for the cumulative observation time. The best-fit double-blackbody model (2BB) is shown in red, with the two blackbody components shown as dashed black and blue lines. The residuals of our best-fit model are shown in the bottom panel.}
    \label{fig:nicer_spec_full}
\end{figure}

Single-model fits prove to be poor matches to the \textit{NICER} data. A single blackbody results in $\chi^2/dof = 644.6/136$ and a single power-law in $\chi^2/dof = 1643.5/136$. Similarly, a double-power-law proves a very bad fit, with strong residuals in both the high- and low-energy ranges. A moderately good fit is achieved with a combination of a blackbody and a power-law, with $\chi^2/dof = 179.3/134$. However, the best fit is found with 2BB, which results in $\chi^2/dof = 144.2/134$, with blackbody temperatures of $k_{\rm B}T_1$ = $105\pm 3$~eV and $k_{\rm B}T_2$ = $59\pm 1$~eV. This best-fit model and the spectral data are shown in Figure~\ref{fig:nicer_spec_full}. Using 2BB, we find the highest flux in the 0.2--2.0 keV light curve to be $1.10\times10^{-11}$ erg~cm$^{-2}$~s$^{-1}$.

To investigate whether the strong changes in count rate over time correlate with spectral changes, we first considered whether there is any evolution in the spectral hardness ratio (HR) over time. We defined the HR using the 0.3--0.5 keV count rate (L) and the 0.5--2 keV count rate (H) as (L-H)/(L+H). Epochs with a count rate below zero were excluded from this calculation. We find no significant evolution in the HR over time. Next, we split the combined \textit{NICER} data up and compare the spectra for the different subsections. Fitting results are shown in Table~\ref{tab:nicer_subsp_fitres}. The first split is by time: we divide the integrating period in two at MJD 59222.80, such that each sub-spectrum has approximately the same number of counts, and we extract the spectrum for the first and the second half. Fitting both spectra with our 2BB model (only redshifts and Galactic $N_\textrm{H}$ frozen), we find no significant differences in the fitting parameters. Secondly, we split our data into periods of high and low count rate. We define the `high state' as those periods where the count rate in the 0.3--1~keV range is greater than 3.78~ct~s$^{-1}$ and the `low state' as the periods where the count rate is below this value. 
We find a slightly higher temperature for one of the blackbodies in the high count spectrum; however, the difference is less than 2$\sigma$.

Stronger evidence for spectral changes was found when we fixed the BB-temperatures to the values found for the \textit{XMM-Newton} spectrum and only let the normalisations vary: the fit for the high-state \textit{NICER} data is considerably better than for the low-state data. Comparing the goodness-of-fit for the models with BB-temperatures fixed to the \textit{XMM-Newton} values with the \textit{NICER} best-fit models (for the high and low state, respectively), we find $\Delta\chi^2 = 23.54$ for the high-state data and $\Delta\chi^2 = 57.02$ for the low state. We note that although the \textit{XMM-Newton} data correspond to a significantly higher flux level than the average of the \textit{NICER} data, the flux level around the peaks in the \textit{NICER} emission is comparable to the \textit{XMM-Newton} level (Figure~\ref{fig:xray_combined}). Next, we compared the high-state and low-state spectra by freezing the value of the BB-temperatures to those in the low-state best-fit model and fitting this model to the high state data. We find a value of $\Delta\chi^2 = 70.76$, compared to the high-state best fit, a significant difference. The changes in X-ray luminosity therefore appear tied to spectral changes. This agrees with the trend visible in the \textit{Swift} XRT data (Table~\ref{tab:xrtfits}), which show higher $k_{\rm B}T_\mathrm{BB}$ for higher flux levels.

\begin{table}
\renewcommand*{\arraystretch}{1.4}
    \centering
    \caption{
    Fitting parameters and goodness-of-fit values for X-ray spectral fits to subsets of \textit{NICER} data.
    }
    \begin{tabular}{l|S[table-format=3.1]cc}
    \toprule
    Sub-spectrum$^a$ & \multicolumn{1}{c}{$k_\textrm{B}T_1^b$} (eV) & $k_\textrm{B}T_2^b$ (eV) & $\chi^2 / dof$\\
    \midrule
      Time: 1$^{st}$ half & 94.5$_{-3.8}^{+4.4}$ & 55.9$_{-2.5}^{+2.3}$ & 192.1/165 \\
      Time: 2$^{nd}$ half & 106.3$_{-4.2}^{+4.5}$ & 59.8$_{-1.7}^{+1.6}$ & 239.2/165 \\
      \midrule
      Rate: low & 89.6$_{-6.4}^{+8.6}$ & 54.7$_{-5.4}^{+4.2}$ & 198.7/165 \\
      Rate: high & 113.6$_{-7.4}^{+8.5}$ & 61.6$_{-2.9}^{+2.5}$ & 184.3/165 \\
    \bottomrule 
    \end{tabular}
    \flushleft{\scriptsize{
    a) The data are divided by time and by count rate as described in the text.\\
    b) Temperatures for fits with a dual-blackbody model.\\
    }}
    \label{tab:nicer_subsp_fitres}
\end{table}

\section{Follow-Up at other wavelengths}
\label{sec:mw}
\subsection{Photometry}
\label{sec:mw_phot}
\begin{figure*}
    \centering
    \includegraphics[width=0.9\linewidth]{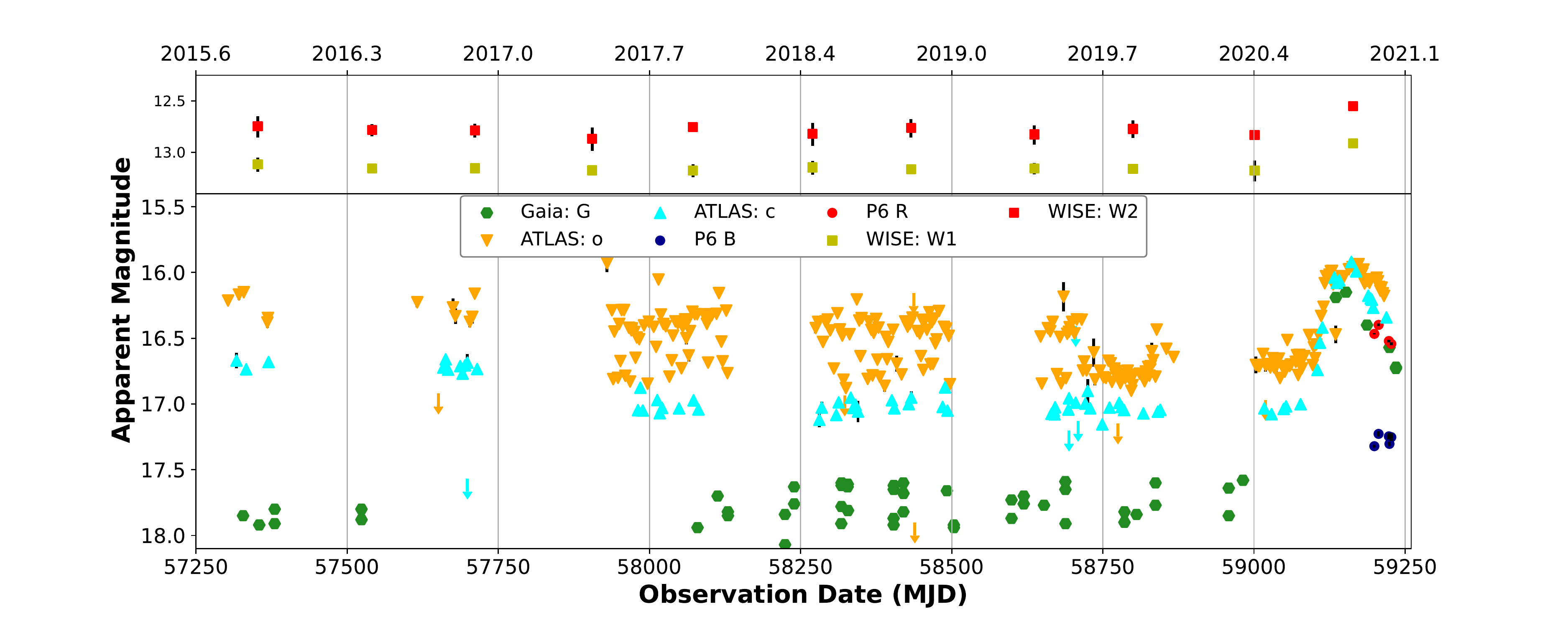}
    \caption{Optical and infrared photometric data on J234402 over the past five years. The top panel show the NEOWISE-R data for the W1 and W2 infrared bands and the bottom panel shows the optical light curve based on a combination of public survey data and follow-up observations obtained by our team. All measurements have been corrected for Galactic extinction. The optical light curve shows a relatively quiescent state interrupted by a sharp flare, followed by a plateau and subsequent decay. The data are not host subtracted. The downward arrows represent 3$\sigma$ upper limits in the ATLAS dataset. All data include statistical uncertainties. For most points on our dataset, the statistical uncertainties are too small to be visible in this plot.}
    \label{fig:opt_lc_full}
\end{figure*}

We present an optical photometric dataset that combines publicly available data and follow-up observations obtained by our team. These data allow us to study the optical evolution of J234402 at high cadence. Optical imaging shows that the ignition is in the northern part of a group of four galaxies (see Figure~\ref{fig:spec_group}). In the following sections we discuss the different datasets making up our photometric data. We correct all our photometry for Galactic reddening using the extinction curve from \citet{FITZPATRICK_1999} and A$_{\rm V}=0.0386$, which we base on the dust map and filter-band corrections provided by \citet{SCHLAFY_2011}\footnote{Retrieved from https://irsa.ipac.caltech.edu/applications/DUST/}. Our combined optical light curve is shown in the lower panel of Figure~\ref{fig:opt_lc_full} and we present our optical dataset in tabular form in Table~\ref{tab:opt_lc}. All data are reported in AB magnitudes \citep[][]{OKE_1974}.

\begin{table}
\def\arraystretch{1.5}
    \centering
    \caption{
    Overview of the optical photometric data obtained for J234402$^a$.
    }
    \begin{tabular}{@{}ccccc@{}}\toprule
        MJD & Survey & Filter & Magnitude$^b$ & Uncertainty\\
        \midrule
        46266.71 &   POSS2 & POSS2-B &  18.17 &   0.03 \\
        50690.58 &   POSS2 & POSS2-R &  16.87 &   0.03 \\
        57303.00 &   ATLAS &       o &  16.21 &   0.02 \\
        57317.00 &   ATLAS &       c &  16.67 &   0.06 \\
        57321.00 &   ATLAS &       o &  16.17 &   0.04 \\
        57327.86 &    Gaia &       G &  17.85 &   0.00 \\
        57329.00 &   ATLAS &       o &  16.15 &   0.02 \\
        57333.00 &   ATLAS &       c &  16.74 &   0.02 \\
    \bottomrule
    \end{tabular}
    \flushleft{\scriptsize{
    a) The full dataset is available in electronic format at the CDS.\\
    b) Not corrected for host galaxy contribution. Upper limits are not included in this table.\\
    }}
    \label{tab:opt_lc}
\end{table}

\subsubsection{Public optical surveys}
We obtained publicly available photometric data from the ATLAS survey. The ATLAS data were taken with filters cyan (420--650 nm) and orange (560-–820 nm) and were extracted by running forced photometry on the available reduced images \citep{TONRY_2018,SMITH_2020}. Observations with S/N$<$3 do not provide a reliable magnitude and are represented as 3$\sigma$ upper limits. We made use of ATLAS difference photometry, in which the pre-ignition flux (dominated by the host galaxy) has been subtracted. As the $o$ and $c$ filters are not included in the filter-specific de-reddening corrections presented by \citet{SCHLAFY_2011}, we made our own estimate of the attenuation for these filters. We simulated a reddened signal, based on a flat spectrum, and convolved this with the filter transmission curves. We took the ratios of the original (non-attenuated) and reddened signals as an approximation of the flux lost to absorption. The corrections applied to the $o$ and $c$ magnitudes in our dataset are 0.027 and 0.039 mag, respectively. To improve the signal-to-noise, the ATLAS light curves were re-binned to combine data from single nights, using a weighted average to find the flux and errors. We subsequently converted to magnitudes. 

For the \textit{Gaia} data \citep{GAIA_2016}, we extracted the G-band magnitudes from the \textit{Gaia} Alerts web server\footnote{http://gsaweb.ast.cam.ac.uk/alerts/home}. The G-band functions as a white light filter ($\sim$300--1100 nm). To create the G-band light curve, photometric measurements are averaged over the period in which \textit{Gaia}'s FOV transits the object, following \textit{Gaia}'s scanning pattern. The photometry made available on the alerts server is based on a preliminary calibration. Using the same procedure as for the ATLAS filters, we find a reddening correction of 0.03 for the \textit{Gaia} G-band.

The \textit{Gaia} and ATLAS light curves show that the ignition in November 2020 is a clear break with the level of output in the preceding years. However, both light curves show a low level of variability ($<0.3$ mag) prior to the ignition, which warrants closer inspection. Of particular interest in the ATLAS light curve are the apparent peaks around MJD 58700 and 58840. We used the unbinned ATLAS data to investigate whether any (near-)contemporaneous ATLAS and \textit{Gaia} data show a similarity in trend around these peaks. There is no clear similarity in the light curves around these times, or indeed elsewhere in the light curve. We therefore investigate the likelihood that the pre-ignition changes are dominated by observational uncertainties. 

In the case of the ATLAS data, we take into account that the observations were made with relatively small, terrestrial telescopes and that variable seeing will therefore lead to changes in the contribution of the host galaxy to the reported magnitudes for the nucleus. The amplitude of the peaks in the ATLAS light curve lies within the range of the seemingly random scatter. Using the ATLAS difference photometry, the two features mentioned above do not appear significant. The \textit{Gaia} G-band measurements are based on PSF photometry \citep{EVANS_2018} and as the observations are not affected by seeing, we expect that changes in host contribution are not a significant contribution to the observed scatter. However, an investigation of DR2 photometry \citep{EVANS_2018,ARENOU_2018} found that spurious outliers caused by nearby sources, as well as by problems with background calibration, are likely to be included in \textit{Gaia}'s epoch photometry. These uncertainties are a particular concern for fainter sources (G$>$17). This means that based on our dataset we cannot make a conclusive statement about the presence of low-level variability prior to the ignition.

\subsubsection{New optical follow-up}
J234402 was monitored with the 0.4m PROMPT6 telescope, operated as part of Skynet \citep[an introduction to this network is provided in][]{MARTIN_2019}, using Johnson B and R filters. These observations provided nine additional epochs to the light curve. The images were reduced in the standard manner (bias correction, flat-fielding) and the fluxes were extracted using 3 arcsecond apertures, using 20--25 arcsecond annuli to establish the background. 

\subsubsection{\textit{XMM-Newton} Optical Monitor}
\label{sec:mw_phot_xmm}
The \textit{XMM-Newton} Optical Monitor (OM) observed J234402 with five exposures, simultaneous with the X-ray observation. Each observation was made both in image and fast mode, using the UVM2 filter (effective wavelength 231 nm). The first two exposures were 1200 s each; the final three were 2500 s each. We reduced the data using \textsc{omichain} and \textsc{omfchain}, which are part of XMM\_SAS. These routines perform flat-fielding, source detection, and aperture photometry for each individual exposure. They also combine all images into a mosaiced image and perform source detection and aperture photometry on the mosaiced image. The resulting values are corrected for detector dead time.
 
We verified that the source was detected in each exposure, confirmed that the source was not too close to the edge of the window in fast mode, and established that there were no obvious imaging artefacts in any image mode exposure. We find that the mean UVM2 OM count rate is 1.65 ct s$^{-1}$, which corresponds to a flux density of $4.23\times10^{-15}$ erg cm$^{-2}$ s$^{-1}$ \AA$^{-1}$ and a magnitude of 16.07. To explore variability down to timescales of 800~s, we made use of the fast mode data. The optical variability is significantly less than the variability of J234402 in X-rays (see Figure~\ref{fig:xmm_lc}). We find $F_\textrm{var}$ of $<1.0\%$ and $20.2 \pm 0.9 \%$, for optical and X-rays, respectively.

\subsubsection{\textit{Swift}-UVOT}
\label{sec:mw_opt_swift}
We observed J234402 with UVOT in four filters in each of the five \textit{Swift} observations: U, UVW1, UVM2, and UVW2 (central wavelengths 3465, 2600, 2246, and 1928 \AA, respectively). Aperture photometry was performed using the ASI SSDC's Multi-Mission Interactive Archive online tool.\footnote{https://www.ssdc.asi.it/mmia/index.php?mission=swiftmastr}. Using this tool we derive aperture-corrected, background-subtracted, and Galactic extinction-corrected flux densities. Source extraction was performed using \textsc{uvotdetect} and CALDB version 20201026. We extracted source counts from a radius of 5 arcsec around the target and the background from an annulus with inner and outer radii of 27.5 and 35 arcsec, respectively. AB magnitudes were calculated using the zero points from \citet{BREEVELD_2011}. We corrected for potential instrumental variability using two standard stars in the FOV (see Appendix~\ref{sec:app_optuv_swift} for further details). The corrected magnitudes and flux densities are listed in Table~\ref{tab:uvotmags}.

The UV brightness in all filters drops over the course of our monitoring, in line with the decreasing X-ray flux. At the time of the first \textit{Swift} observation, the optical fluxes were also in decline. Over the course of the \textit{Swift} observations, covering approximately 33 days, the UV fluxes decrease by an average of $\sim$0.35.  The decrease in flux was more pronounced in the X-rays as the 0.2--2~keV flux dropped $\sim$0.45 over the same period. The UV fluxes have not been corrected for the contribution from the host galaxy; however, we expect this contribution to be stable on a timescale of weeks. In all five observations, the UVOT magnitudes are considerably higher than the archival UV values we find in the Revised Catalogue of GALEX UV Sources \citep[][]{BIANCHI_2017}. In the NUV and FUV bands, with effective wavelengths of 1516 \AA\ and 2267 \AA, the source was observed in 2005 as part of the all-sky imaging survey at $19.94\pm0.07$ and $20.83\pm0.16$ mags, respectively. This is consistent with the archival limits we find in the X-rays and optical bands: J234402 was significantly less luminous in the years prior to the ignition event.

\hspace{-5cm}
\begin{table*}\centering
\def\arraystretch{1.5}
\caption{
\textit{Swift}-UVOT magnitudes for J234402.
}
    \begin{tabular}{lcccc}
        \toprule
        Obs$^a$ &  U$^b$ mag &  W1 mag &  M2 mag &  W2 mag \\ 
        \midrule
         59203.98   & $15.53\pm0.05$ & $14.90\pm0.05$ & $15.05\pm0.05$ & $14.75\pm0.04$ \\
         59210.29   & $15.75\pm0.04$ & $15.15\pm0.05$ & $14.97\pm0.05$ & $14.84\pm0.04$ \\
         59220.38   & $15.66\pm0.04$ & $15.14\pm0.05$ & $14.92\pm0.05$ & $14.88\pm0.04$ \\
         59232.07   & $15.74\pm0.05$ & $15.18\pm0.06$ & $15.10\pm0.06$ & $15.03\pm0.05$ \\
         59236.51   & $15.92\pm0.05$ & $15.39\pm0.06$ & $15.16\pm0.06$ & $15.15\pm0.04$ \\
    \bottomrule
    \end{tabular}
    \flushleft{\scriptsize{
    a) The values have been aperture-corrected and adjusted for Galactic extinction. We have also calibrated for possible instrumental offsets, using standard stars in the UVOT FOV.\\
    b) The relevant UVOT-filter central wavelengths are 3465, 2600, 2246, and 1928 \AA\ for U, W1, M2, and W2, respectively.\\
    }}
\label{tab:uvotmags}
\end{table*}

\subsubsection{\textit{WISE}}
\label{sec:mw_phot_wise}
We obtain infrared photometry from the NEOWISE-R mission \citep{WRIGHT_2010,MAINZER_2014}, using the NASA IRSA archives. The WISE bands W1 and W2 are centred on 3.368 and 4.618 microns, respectively. We use the fluxes generated by the automated forced photometry pipeline. WISE clearly resolves the northern and southern sources. Coordinate matching for the NEOWISE-R extraction was limited to a 5 arcsec radius. The fluxes are binned to $\sim$180 days, to match the satellite's scanning pattern, and converted to magnitudes. We applied a sigma-clip to the flux measurements in each bin to remove a small number of clearly erroneous flux entries, after which we calculated the error-weighted mean flux. The resulting magnitudes are included in the top panel of Figure~\ref{fig:opt_lc_full}.

The WISE light curves show a distinct brightening, by approximately 0.3 magnitude in W1 and W2, around the time of the ignition event. We investigated the significance of this change by calculating the standard deviation of the fluxes prior to the ignition event and comparing this value to the size of the increase. We find that the standard deviation is $68.8\pm19.6$ for the W1 band and $57.1\pm20.1$ for the W2 band. The fluxes are in raw counts, following background subtraction, and we used a bootstrap Monte Carlo method to estimate the uncertainties. 
Compared to the average flux over the pre-ignition period, the flux increased by 280 counts and 131 counts around the time of ignition for W1 and W2, respectively, representing changes by 4.1$\sigma$ and 2.3$\sigma$. We therefore believe that the increase in infrared flux is not associated with the `normal' stochastic variability exhibited by J234402 prior to the ignition event, but rather that the increase in IR is connected to the increases in X-ray, UV, and the optical. 

\subsubsection{Light curve properties}
\label{sec:mw_opt_lc}
\begin{figure}
    \centering
    \includegraphics[width=0.5\textwidth]{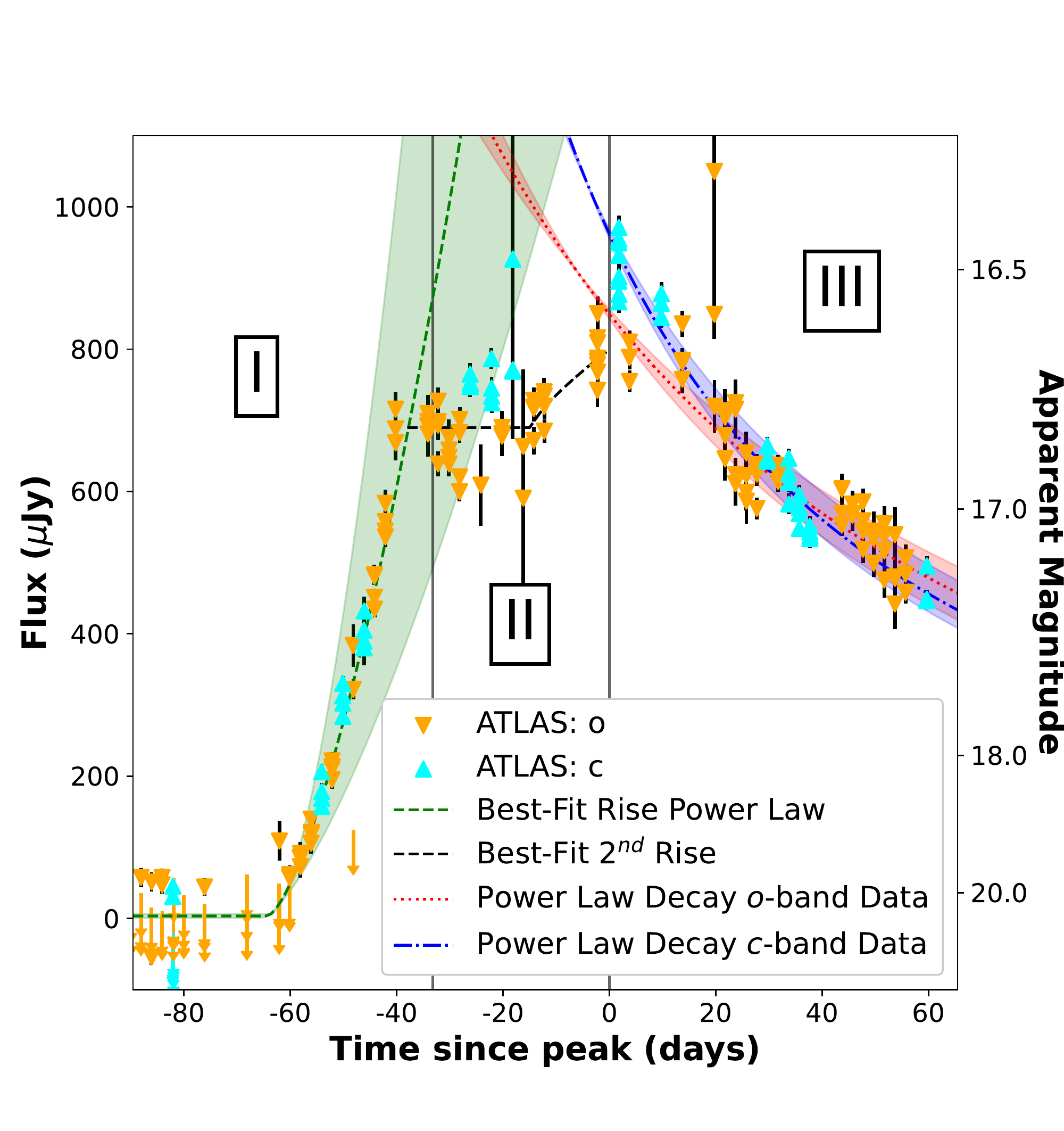}
    \caption{
    Evolution of optical output in ATLAS $o$-band and $c$-band filters. The data are ATLAS differential photometry and the view focuses on the time around the ignition event. All data points include statistical errors. We distinguish three phases in the first months of the ignition event's light curve: a sharp initial rise, a temporary levelling off \& re-brightening, and decay. The demarcations between the phases are shown as grey vertical lines. We compare the data in these three phases with simple power-law models (Equations~\ref{eq:rise} and~\ref{eq:decay}). The initial rise is fit to the $o$-band data only and the result is shown in dashed green, with 1$\sigma$ error margins shown as shaded regions. We include the best-fit rising power-law in phase II (dashed black) primarily to guide the eye, as the light curve data proved to sparse for a well-constrained fit. For the decay period, we separately fit the $o$-band and the $c$-band data, and results are shown as the red (dotted) and blue (dot-dashed) lines, respectively.
    }
    \label{fig:lc_fitting}
\end{figure}

We made use of the ATLAS difference photometry to find the time of peak and to constrain the slope of the rise. As the pre-ignition flux has been subtracted, we consider the difference photometry to represent the flux of the brightened nucleus only. A detailed view of the host-corrected ATLAS data around the time of the ignition is shown in Figure~\ref{fig:lc_fitting}. We mark three distinct phases in the light curve: I)~the sudden turnover from the quiescent state into a sharp rise; II)~a temporary flattening in the brightening, lasting approximately 30 days, followed by a renewed increase in flux leading to the peak output; III)~decay after the peak, lasting until the period of Sun block. We fitted the data in the three phases separately and derived key parameters that describe the early evolution of J234402's light curve.

First we established the point of turnover, where the initial rise halts temporarily. As the sampling of the $c$-band data is too sparse over this period, we performed this fit for the $o$-band data only. We fitted the flux data with a parabola, matching the rapid rise, quick turnover, and the apparent temporary decrease in flux. As the halt in the brightening is quite brief, we fitted our parabola in a relatively narrow range, similar to the procedure in for example \citet{HOLOIEN_2022} and \citet{HINKLE_2022}, fitting between MJD 59111 and 59129. We estimated the error on our fit using a Monte Carlo method. We created $10^4$ iterations of our light curve, where the flux in each bin was shifted by a random value drawn from a Gaussian distribution with $\sigma$ equal to the measurement uncertainty of the flux. We fitted each of these iterations and took the median values to be our best-fit parameters. We took the 16th and 84th percentiles of the parameter distributions as the uncertainties. Using this method we find $t_{\rm peak,1}$ = MJD $59126.31^{+0.52}_{-0.45}$. This first `peak' is marked with the left grey, vertical line in Figure~\ref{fig:lc_fitting}. It forms the delineation between Phase~I and Phase~II.

Next, we fitted the early-time rise of the light curve using all $o$-band data up to $t_{\rm peak,1}$. We modelled the flux as constant prior to the ignition at $t_{\rm start}$ and as a power law afterwards:
\begin{equation}
\label{eq:rise}
  f(t) =
  \begin{cases}
          h & \text{if $t < t_{\rm start}$} \\
          h + A(t-t_{\rm start})^{b} & \text{if $t\geq t_{\rm start}$,} 
  \end{cases}
\end{equation}
where $t$ is in days and the free parameters in our least-squares fit are the constant level $h$, the normalisation $A$, and the power-law index $b$. The uncertainties were estimated in the same manner as we did for $t_{peak,1}$. We find $t_{\rm start} = \textrm{ MJD } 59095.31_{-0.93}^{+0.82}$, $h=3.49_{-2.97}^{+2.93}$ $\mu$Jy, $A=5.30_{-1.50}^{+1.83}$ $\mu$Jy, and the power-law index $b = 1.48_{-0.08}^{+0.09}$. The best fit is shown in green (dashed) in Figure~\ref{fig:lc_fitting}, as well as the 16th and 84th-percentile uncertainties, indicated by the shaded regions. 
Comparing t$_{\rm peak,1}$ with $t_{\rm start}$, we find a total rise time of $31.00_{-1.03}^{+0.94}$ days. The rise period is within the range of values found for various nuclear transients: the TDE ASASSN-19bt \citep[$41.2\pm0.5$ days ;][]{HOLOIEN_2019}, the TDE ASASSN-19dj \citep[$\sim$ 27 days;][]{HINKLE_202108}, the TDE AT2019qiz \citep[30.6 days;][]{NICHOLL_2020}, and the ambiguous nuclear transient ASASSN-20hx \citep[$\sim$29 days;][]{HINKLE_2022}. The power-law index $b$ is lower than a quadratic rise, which was found for ASASSN-19bt and  ASASSN-19dj.

To estimate the time of the main peak in the light curve we again fitted a parabola, this time in the range MJD 59127--59191. We find $t_{\rm peak,2}$ = MJD $59159.56^{+0.38}_{-0.37}$. This value is used as the peak time in Figure~\ref{fig:lc_fitting}. The $o$-band light curve between $t_{\rm peak,1}$ and $t_{\rm peak,2}$, Phase II in our schema, shows a plateau, followed by a rise to the main peak. We fitted the data with the function defined in Equation~\ref{eq:rise} and find a shallower increase, with $b = 0.71_{-0.13}^{+0.20}$ (the black dashed line in Figure~\ref{fig:lc_fitting}).

Finally, we fitted the decay in Phase III. While the light curve of a TDE is canonically expected to decline post-peak according to a $t^{-5/3}$ power-law profile \citep[i.e. in the fallback-dominated regime;][]{REES_1988,PHINNEY_1989}, the observed values are known to have a wide range. We used the following model based on the decay model used by \citet{VANVELZEN_2021}:
\begin{equation}
\label{eq:decay}
    f(t) = C \left( \frac{t-t_0}{t_{sc}}+1 \right) ^p \hspace{5mm} \textrm{if $t>t_0$}.
\end{equation}
We separately fitted the decay function to the $o$-band and the $c$-band data. We fixed $t_0$ to $t_{peak,2}$ and allowed the other parameters to vary freely. For the $o$-band, we find time-scaling factor $t_{sc} = 186.45_{-12.87}^{+16.20}$ days, normalisation $C = 848.97_{-6.91}^{+6.90}$ $\mu$Jy, and power-law exponent $p = -2.06_{-0.11}^{+0.09}$. The $c$-band fluxes appear to have a slightly steeper decay, as we find $t_{sc} = 80.28_{-3.86}^{+4.14}$ days, $C = 971.94_{-6.05}^{+6.01}$ $\mu$Jy, and $p = -1.34_{-0.04}^{+0.04}$. Although the exponent is less negative for the bluer photometry band, $t_{sc}$ is significantly smaller. As $t_{sc}$ pushes the start of the decay period back in time, a large value of $t_{sc}$ represents a shallower decay. We note that the freedom in setting both $t_{sc}$ and $p$ means that our function can be used to model a wide range of light curves that show any form of decay over time.

\subsection{Spectroscopy}\label{sec:mw_spec}
We obtained several long-slit spectra of the optical counterpart of J234402 as well as of the three close-by southern sources. No archival spectroscopic observations exist of J234402 prior to the outburst, although a 2dFGRS \citep{COLLESS_2003} was taken of the blend of the southern sources (Appendix~\ref{sec:app_optuv}).

\begin{figure*}
    \centering
    \includegraphics[width=.65\linewidth]{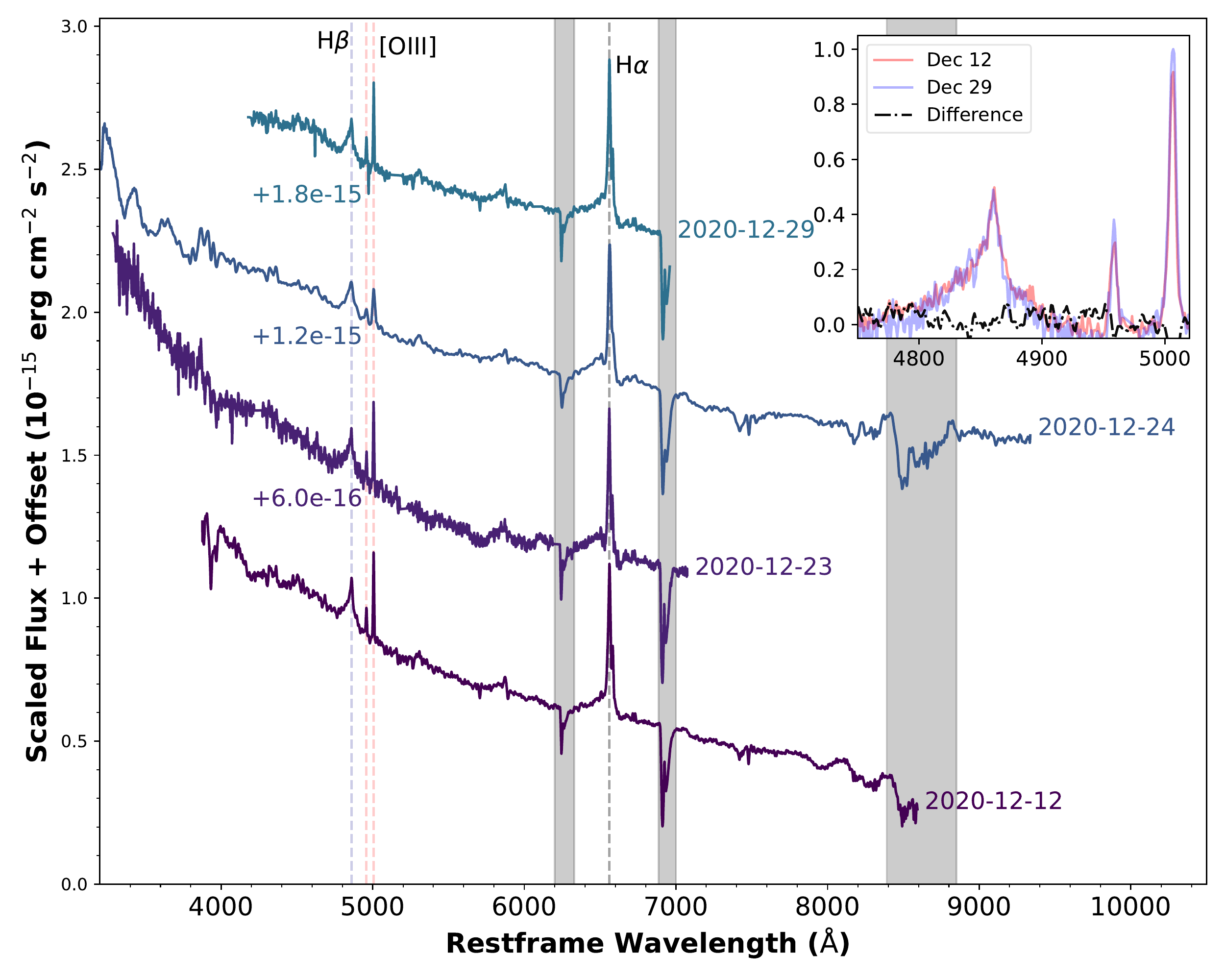}
    \caption{Optical spectra of J234402. The data are from our follow-up campaign with Baade (Magellan), VLT, and SALT. The spectra show a blue continuum as well as strong broad and narrow emission lines. The broad lines show a distinct asymmetry, with a blue wing. The spectra have been scaled using the P6 photometry and are presented here with a flux offset for clarity. Several important emission lines are marked with dashed lines. Spectral regions affected by telluric absorption have been indicated by grey bands. The inset shows a close-up of H$\beta$ for the two spectra separated by the largest time span (17 days), as well as the difference between the spectra. The fitted host+continuum components have been subtracted from the spectra in the inset and they were normalised using the [OIII]$\lambda$5007 line fluxes. Over the period of follow-up presented here, the object shows no significant spectral evolution.}
    \label{fig:opt_spec}
\end{figure*}

Our new observations were made from the beginning of December 2020 until the Sun-block period (January 2021). The spectra were taken using the IMACS Short-Camera \citep[][]{DRESSLER_2011} mounted on the 6.5m Baade Magellan telescope located at Las Campanas Observatory, the Robert Stobie Spectrograph \citep[RSS;][]{BURGH_2003,KOBULNICKY_2003} mounted on the 10m South African Large Telescope \citep[SALT;][]{BUCKLEY_2006}, and the FORS2 instrument \citep{APPENZELLER_1998} mounted on the 8.2m Very Large Telescope Array's UT1 at Cerro Paranal. An overview of the observations is listed in Table~\ref{tab:opt_spectra}. All four observations covered J234402, and the Baade and SALT observations together covered the three southern sources. Our spectra of J234402 are shown in Figure~\ref{fig:opt_spec}. The spectra were taken over a relatively brief interval (17 days), prior to Sun block and we observe no significant spectral changes over this period. We note, however, that in our follow-up after the Sun-block period, we do observe considerable spectral evolution, thus strengthening the case for the association of the X-ray flare with the optical outburst. We will discuss these later spectra in a follow-up paper.

The IMACS and RSS observations were taken at non-parallactic angles, to cover the southern sources. As both SALT and Baade make use of an Atmospheric Dispersion Corrector, the non-parallactic observations had no impact on the observed spectral shapes. Using the same parameters as for our photometry, all spectra were corrected for Galactic extinction. We scaled the absolute flux level of the spectra using contemporary photometry. For the spectrum taken on 23 December with SALT, we have R-band photometry from P6 taken on the same night. We scale the SALT spectrum so that the derived R-band flux matches that of our photometry. Subsequently we fitted the narrow [OIII]$\lambda\lambda$4959,5007 lines in all four spectra (using a single Gaussian for each line). We scaled the other three spectra to the photometry-scaled SALT spectrum, by matching the integrated flux of the [OIII] lines.
\begin{table}
\def\arraystretch{1.5}
    \centering
    \caption{Overview of our optical spectroscopic observations of J234402.}
    \begin{tabular}{@{}ccccc@{}}\toprule
        MJD & Instrument (Telescope) & Slit width & Seeing \\
        \midrule
        59195 & IMACS (Baade) & 0.7" & 0.65" & \\
        59206 & RSS (SALT) & 1.5" &  1.2"\\
        59207 & FORS2 (VLT) & 3" & 0.35--1.63" \\
        59212 & RSS (SALT) & 1.5" & 1.3" \\
    \bottomrule
    \end{tabular}
    \label{tab:opt_spectra}
\end{table}

\subsubsection{Spectroscopic analysis}
\label{sec:mw_spec_fit}
\begin{figure*}[h!]
    \centering
    \includegraphics[width=0.78\linewidth,trim={2cm 2cm 1cm 3cm},clip]{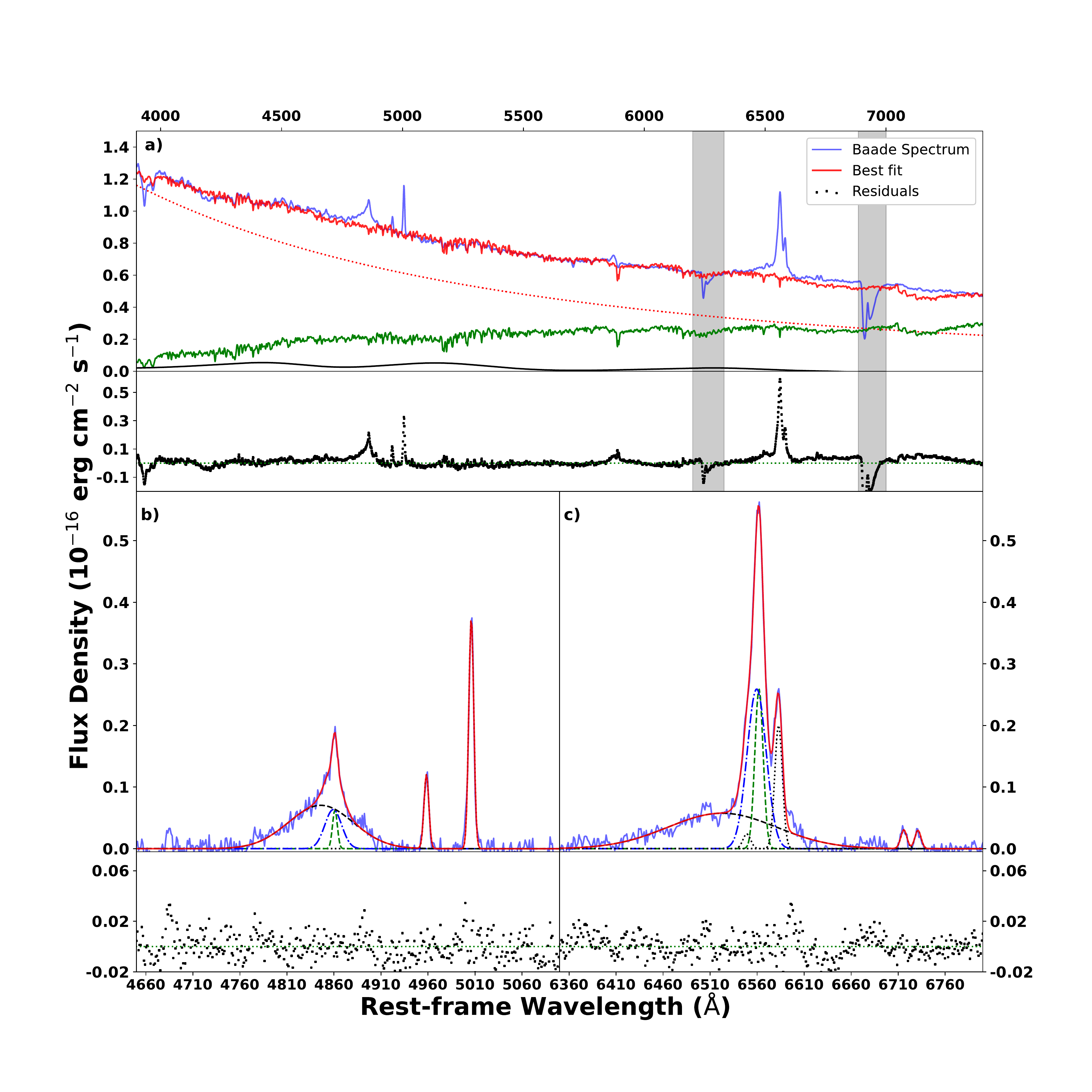}
    \caption{Our fitting procedure for the optical spectra. The procedure is illustrated for the spectrum taken with the Baade telescope on MJD 59195. \textit{a)} Continuum fit. The continuum is fit with a function consisting of a single power law (\textit{red dotted}), an FeII-emission template (\textit{black}), and a host-galaxy template (\textit{green}). The residuals for all fits are shown underneath their respective plots. \textit{b)} Fit of continuum-subtracted spectrum around H$\beta$. The combined fit consists of five Gaussian profiles, two for the narrow [OIII] (\textit{black dotted}) lines and a narrow (\textit{green dashed}), broad (\textit{blue dot-dashed}), and very broad (\textit{black dashed}) component for H$\beta$. \textit{c)} Fit of continuum-subtracted spectrum around H$\alpha$. The combined fit for this region consists of six Gaussians: two for the narrow [NII] lines, two for the narrow [SII] lines (\textit{dotted black}), and three for the H$\alpha$ emission, with the same colours as the components used to fit H$\beta$.}
    \label{fig:opt_fit_lines}
\end{figure*}

The spectra of J234402 show a blue optical continuum and strong, asymmetric Balmer emission lines. To analyse the spectra, we performed a least-squares fit to the data, making use of the \textit{lmfit} package\footnote{https://lmfit.github.io/lmfit-py/}. The fitting process was iterative: the first step was to fit a model consisting of a power-law continuum, an FeII template, and a host-galaxy template to the data. The FeII-template is the empirical (1 Zw I-based) template presented in \citet{BRUHWEILER_2008}. For the host template we made use of a spectrum produced from the stellar population synthesis models of \citet{BRUZUAL_2003}, using a model of 11.5 Gy and 0.05 solar metallicity. The strength of the host contribution was established in the fit to the Baade spectrum, after which this parameter was fixed in the fitting of the other three spectra. The second step was to fit the residuals of the continuum+FeII+host fit with Gaussians, to approximate the emission line profiles. The fitting results for the Baade spectrum of J234402, the first spectrum taken after the eROSITA detection, are shown in Figure~\ref{fig:opt_fit_lines}. We list the average fitting parameters in Table~\ref{tab:opt_fit_results}.

The slope of the continuum power law (f$_{\lambda}\propto\lambda^{-\alpha}$) is $\sim$2.5, a value typical of luminous AGN \citep[e.g.][]{VANDENBERK_2001}. There is no indication of a strong FeII contribution to the spectrum, and we detect no individual FeII emission lines. The FORS2 spectrum has features below 4000 \AA\, however these are most likely not physical (see Appendix~\ref{sec:app_optuv} for a detailed discussion). To fit the regions around H$\alpha$ and H$\beta$, we made use of a locally fitted continuum. The fitting regions are 6100--7000~\AA\ and 4600--5200~{\AA}, for H$\alpha$ and H$\beta$, respectively. To account for the strong asymmetry in the lines, we used three Gaussian components (narrow, broad, and very broad). We find that both H$\alpha$ and H$\beta$ include a strong, very-broad component, which is offset to the blue compared to the two narrower components (Figure~\ref{fig:opt_fit_lines}). In both Balmer lines, the broadest component is quite distinct from its narrower counterparts. The very-broad components have full widths at half maximum (FWHM) of $\sim$5000 km~s$^{-1}$, significantly wider than the broad components. The offsets between the centres of the broad line components and the narrow line components are 52.4$\pm$1.3 \AA\ and 12.6$\pm$1.3 \AA\ for H$\alpha$ and H$\beta$, respectively. If we associate this Doppler shift with an inflow or outflow, the associated radial flow speeds would be of the order of 1,000 km s$^{-1}$. 

The Balmer decrement, f(H$\alpha$)/f(H$\beta)$, for the very broad components is $\sim$1.4, which remarkably is below the expected recombination value of approximately 3. We checked the impact of our fitting method on these results by refitting H$\beta$ in our highest S/N spectrum (Baade), but limiting the flux in the very broad component to 1/3*f(H$\alpha_{VB}$). This resulted in an acceptable fit, where the broad-line model component compensates for the lowered very-broad component. However, in this case the Balmer decrement for the broad component drops below 3. In fact, we find that the Balmer decrement calculated using the sum of the broad and very broad Gaussians remains approximately constant at 2.1, under various constraints. As the continuum subtraction is of considerable importance in estimating the line fluxes, we experimented with various fits to the local continuum around the lines, but find no acceptable solutions that would result in a Balmer decrement $\geq$3. We therefore conclude that this unusual flux ratio is intrinsic to the spectrum of J234402, although its origin remains unclear. Interestingly, \citet{LI_2022} find a low value of the Balmer decrement immediately following the flare in 1ES 1927+654, as low as $\sim$1, which increases to $>$3 on a timescale of hundreds of days. We will continue to monitor the evolution of this anomalous line ratio in J234402, to investigate whether it is related to the rapid changes in the BLR.

Two Helium lines are visible in the residuals of our spectral fitting (Figure~\ref{fig:opt_fit_lines}). He~I~$\lambda$5876 is clearly detected: it is mixed with an absorption feature, which we associate with the Na~I D1 and D2 doublet ($\lambda\lambda$5889.9, 5895.9). Due to this blend of features we did not fit the line profile for He~I~$\lambda$5876. We do note that there is no visible change in this line among our spectra. He~II~$\lambda$4686 is only detected as a narrow emission line. This is somewhat surprising, as the strong soft X-ray emission could be expected to be associated with a strong ionising continuum, powering strong broad He~II emission. A significant mismatch among line strengths, although unusual, has been detected in AGN samples \citep{FERLAND_2020}, in individual highly variable AGN over time \citep[e.g.][]{HOMAN_2023}, and among TDEs \citep{VANVELZEN_2021}. We calculated an upper limit on the total He~II~flux by summing the flux in the continuum-subtracted spectrum (this therefore includes any narrow line emission) and find 1.4 erg cm$^{-2}$ s$^{-1}$. This results in an upper limit of 2\% for the flux ratio He~II~$\lambda$4686/H$\beta$.

\begin{table}
\def\arraystretch{1.5}
    \centering
    \caption{
    Results of fitting the optical spectra of J234402.
    }
    \begin{tabular}{lrr}
    \toprule
    Model Parameter$^a$ & Value$^b$ & 1$\sigma$ \\
    \midrule
    \textit{Line Flux} & \multicolumn{2}{c}{[10$^{-16}$ erg~cm$^{-2}$~s$^{-1}$]} \\
    \midrule
        f(H$\beta_{VB}$) & 56.4 & 1.8 \\
        f(H$\beta_{B}$) & 12.8 & 1.4 \\
        f(H$\beta_{N}$) & 4.6 & 0.7 \\
        f(H$\alpha_{VB}$) & 80.2 & 1.5\\
        f(H$\alpha_{B}$) & 64.4 & 1.4\\
        f(H$\alpha_{N}$) & 28.3 & 0.8\\
        f([OIII]$\lambda$5007) & 24.6 & 0.3 \\
        f([NII]$\lambda$6585) & 19.0 & 0.4 \\ 
        f([SII]$\lambda$6716) & 2.5 & 0.2\\
    \midrule
    \textit{Line Width} & \multicolumn{2}{c}{[km s$^{-1}$]} \\
    \midrule
        FHWM(H$\beta_{VB}$) &  4684 & 141\\
        FHWM(H$\beta_{B}$) &  1561 & 137\\
        FHWM(H$\beta_{N}$) &  406 & 39\\
        FWHM(H$\alpha_{VB}$) & 5942 & 119\\
        FWHM(H$\alpha_{B}$) & 980 & 15\\
        FWHM(H$\alpha_{N}$) & 465 & 23\\
        FWHM([OIII]$\lambda$5007) & 169 & 3 \\
    \midrule
    \textit{Continuum Slope} & \multicolumn{2}{c}{[$-$]} \\
    \midrule
        $\alpha$ & 2.52 & 0.01\\        
    \bottomrule
    \end{tabular}
    \flushleft{\scriptsize{
    a) The parameter $\alpha$ represents the slope of the power-law, f$_{\lambda}\propto \lambda^{-\alpha}$, fit to the continuum over the full wavelength range available in each spectrum.\\
    b) All values represent the weighted average for fits to the four individual spectra.\\
    }}
    \label{tab:opt_fit_results}
\end{table}

\begin{figure*}[h!]
    \centering
    \includegraphics[scale=0.35]{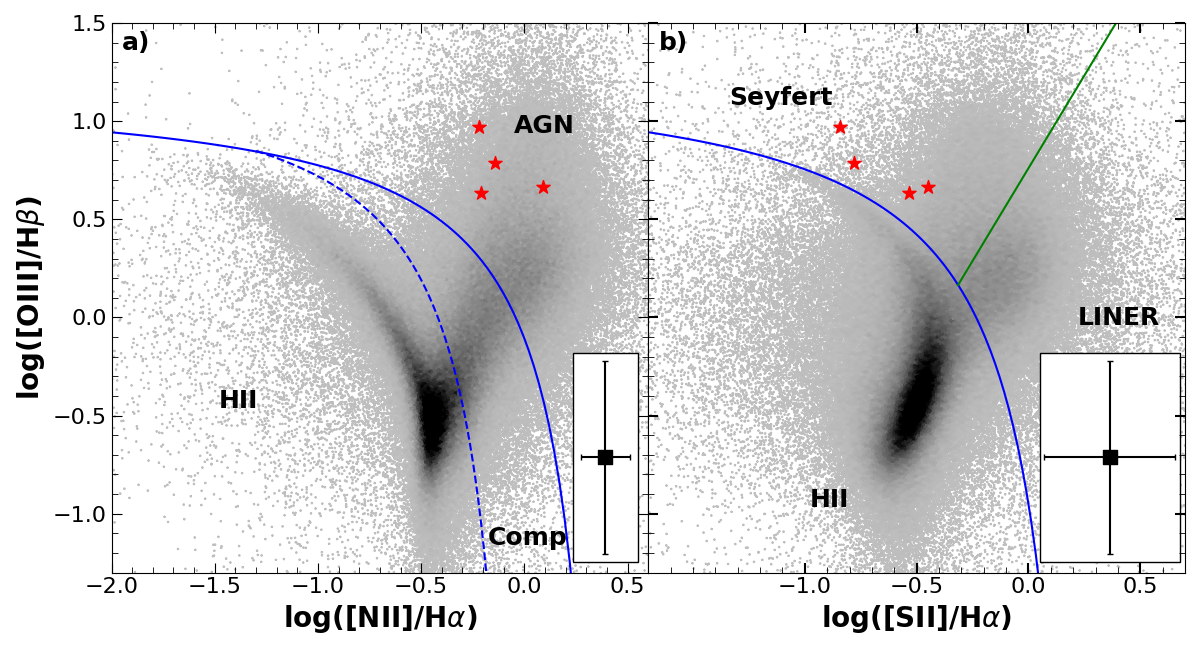}
    \caption{Line flux diagnostic diagrams \citep[][]{BALDWIN_1981} for J234402. The classification curves are from \citet{KEWLEY_2006}. 
    The location of each of the J234402 spectra is indicated with a red star. The two subplots show the ratio [OIII]$\lambda$5007/H$\beta$ $a)$ plotted against [NII]$\lambda$6584/H$\alpha$ and $b)$ plotted against [SII]$\lambda$6717/H$\alpha$. The shaded regions represent the line flux ratios of $\sim$100,000 SDSS galaxies (restricted to $0.05<z<0.15$), using the fluxes made available in the MPA-JHU data release (http://wwwmpa.mpa-garching.mpg.de/SDSS/DR7/). Darker shades of grey mean a higher number density of objects. For the Balmer lines of J234402, the fluxes were calculated using only the narrow components of the line decomposition (Figure~\ref{fig:opt_fit_lines}). The narrow lines are formed at a kpc-scale distance from the central source and therefore represent a past accretion state. The line diagnostics suggest that the nucleus of J234402 was active for an extended period within the last $\sim$10$^3$ years. The inset in the bottom left of each plot shows the mean of the formal uncertainties associated with the J234402 spectral fits.}
    \label{fig:OPT_bpt}
\end{figure*}

We used the line parameters of J234402 to create a diagnostic diagram following \citet[BPT;][]{BALDWIN_1981}. In Figure~\ref{fig:OPT_bpt}, we present the diagrams for [OIII]$\lambda$5007/H$\beta$ compared to [NII]$\lambda$6584/H$\alpha$ and [SII]$\lambda$6717/H$\alpha$. The flux ratios for J234402 fall within the AGN classification of the diagram. As [OI]$\lambda$6100 lies within a telluric absorption region, we cannot provide a reliable estimate using this line. We only used the narrow-line fluxes from our spectral decomposition. In Figure~\ref{fig:OPT_bpt}, we include individual results from each of our spectra in order to highlight the uncertainties involved. The line ratios show the relative strength of the high-ionisation [OIII] lines, which likely require an AGN to power them. Based on consideration of the light-travel-time to the narrow-line region, we can therefore say that the SMBH in J234402 was likely actively accreting for extended periods within the last millennia.

\subsection{\textit{Fermi}-LAT}\label{sec:mw_fermi}
To search for possible $\gamma$-ray emission related to the formation of a jet during the ignition event, we used data from the \textit{Fermi} Large Area Telescope (LAT). \textit{Fermi}-LAT continuously monitors the sky in an energy range from 30\,MeV up to more than 300\,GeV \citep{2009ApJ...697.1071A}. For the data analysis, we used the recent Pass 8 data, including the post-launch instrument response function, \texttt{P8R3\_SOURCE\_V2}, together with the \texttt{ScienceTools} version 1.2.23 and fermipy version 0.20.0. We selected photons with energies between 100\,MeV and 300\,GeV and located within a region of interest (ROI) of $10^{\circ}$ centred on the eROSITA coordinates. We excluded photons that entered the LAT with a zenith angle $>90^{\circ}$. We evaluated the $\gamma$-ray data during two time ranges, which approximately represent the pre-ignition and post-ignition period for J234402. We define pre-ignition as the period from eRASS 1 (24 May 2020) to the \textit{Gaia} transient alert (14 October 2020) and post-ignition as the period from the \textit{Gaia} alert to 31 January 2021, the start of the Sun-block period. 

The data were modelled using a maximum likelihood analysis. The significance of each modelled $\gamma$-ray signal within the ROI is given by the test statistic value $\mathrm{TS}=2\Delta \mathrm{log}(\mathcal{L})$ \citep{1996ApJ...461..396M}. The TS value can be roughly translated into a significance as $\sigma$=$\sqrt{TS}$. Our model consists of known $\gamma$-ray sources from the \textit{Fermi}-LAT Fourth source catalogue \citep[4FGL;][]{2020ApJS..247...33A} within $15^{\circ}$ of the ignition coordinates, the Galactic diffuse model \texttt{gll\_iem\_v07}, and the model for the isotropic diffusion emission \texttt{iso\_P8R3\_SOURCE\_V2\_v1}. The closest 4FGL source has a distance of $2.8^{\circ}$ to the coordinates of the transient; hence no known $\gamma$-ray source can be associated with J234402.

We next attempted to establish the presence of a point source that is not included in the catalogue, by adding it at the counterpart coordinates to our model. We assumed a power-law spectrum with a photon index of 2. After the first fit, we excluded all sources from the model that have $\mathrm{TS}<4$ ($\sigma<2$) or $\mathrm{TS}=nan$. We kept the galactic and isotropic diffuse models free, as well as all parameters and normalisations for sources within $3^{\circ}$. For both the pre-ignition and post-ignition periods, we find no significant detection at the location of J234402. For the pre-ignition period, we find a 2$\sigma$ upper limit\footnote{It is standard for \textit{Fermi}-LAT data to give upper limits at 2$\sigma$ confidence.} of $2.17\times10^{-9}$ photons\,cm$^{-2}$\,s$^{-1}$ ($1.6\times10^{-12}$ erg\,cm$^{-2}$\,s$^{-1}$), and for the post-ignition period, a 2$\sigma$ upper limit of $1.8\times10^{-8}$ photons\,cm$^{-2}$\,s$^{-1}$ ($6.3\times10^{-12}$ erg\,cm$^{-2}$\,s$^{-1}$). We note that the slightly lower upper limit for the pre-ignition period can be explained by the longer time range included in the analysis ($\sim$5 months vs. 3.5 months), and does not necessarily imply a higher $\gamma$-ray flux.

\section{Discussion}
\label{sec:discussion}
\begin{figure*}[h!]
    \centering
    \includegraphics[width=0.65\linewidth]{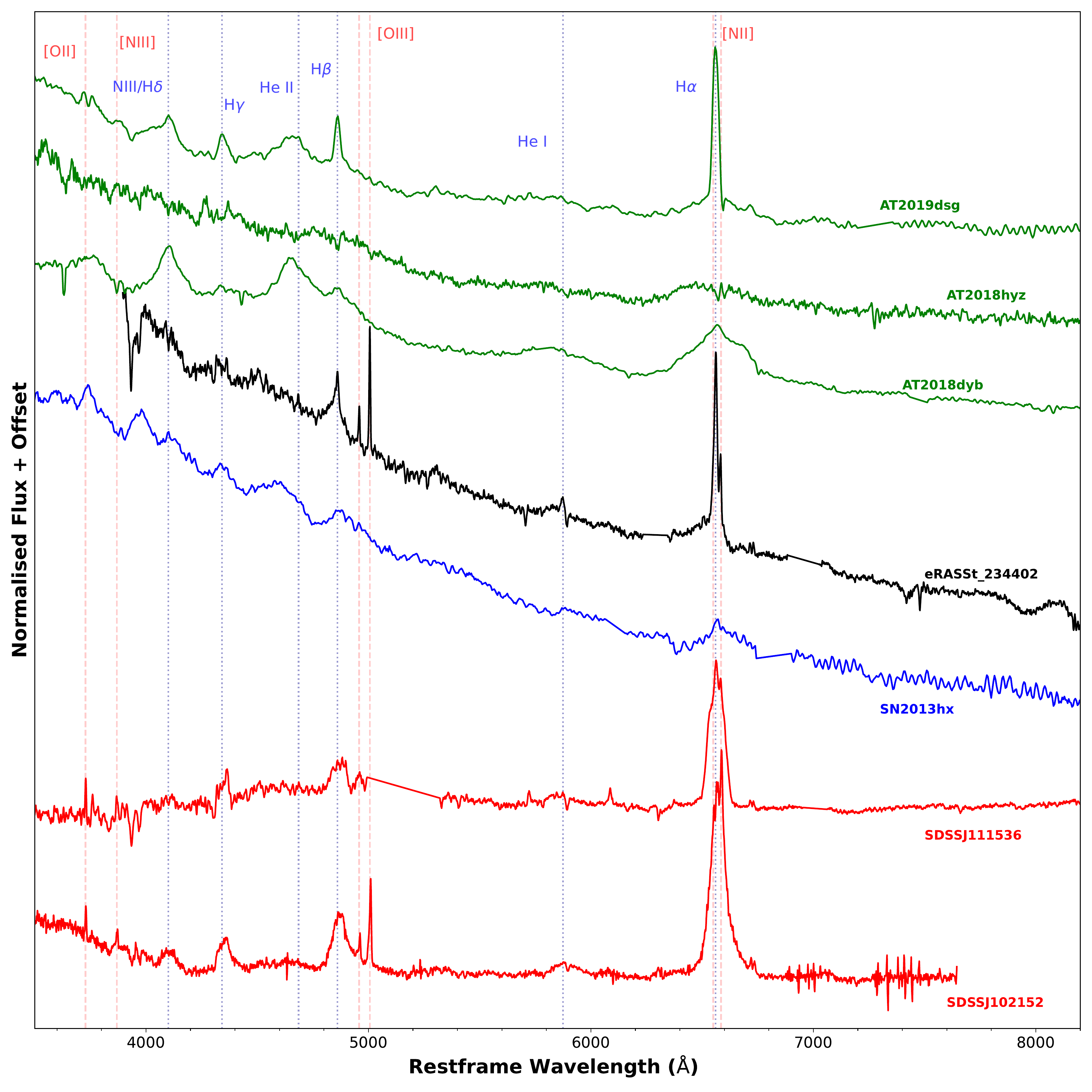}
    \caption{Spectrum of J234402 compared with spectra of several classes of objects. The optical spectrum of J234402 ($black$) is shown together with spectra from several other classes of objects that show extreme outbursts: TDEs ($green$), supernovae ($blue$), and changing-look AGN ($red$). The emission lines marked in red (dashed) are those more typically associated with AGN and the emission lines marked in blue (dotted), are those that are prominent in both AGN and TDEs. The narrow-line features of J234402 are more typical of an AGN; however, the broad-line features agree with both a TDE and an AGN interpretation. The included TDE spectra belong to the TDE--H+He AT2019dsg \citep[$z=0.051$, spectrum 34 days after discovery,][]{NICHOLL_2019}, the TDE--H AT2018hyz \citep[$z=0.046$, spectrum 6 days after discovery,][]{DONG_2018}, and the TDE--H+He AT2018dyb \citep[$z=0.018$, spectrum 34 days after peak emission,][]{LELOUDAS_2019}. The SN spectrum belongs to the H-rich SLSN II SN2013hx \citep[$z=0.125$, spectrum 55 days after detection,][]{INSERRA_2018}. The two changing-look AGN spectra both represent the relative high state of the objects: SDSSJ111536.57+054449.7 \citep[$z=0.090$, first reported by][]{YAN_2019} and  SDSSJ102152.34+464515.6 \citep[$z=0.204$, first reported by][]{MACLEOD_2016}. All spectra have been corrected for redshift, the flux densities have been normalised for the purpose of comparison, and we have removed features that were obviously associated with atmospheric absorption. No correction was made for the host galaxy contribution.}
    \label{fig:DIS_comp_spec}
\end{figure*}

J234402 is a strong extragalactic ignition event with observed changes in X-rays, UV, optical emission, and IR. We aim to place J234402 in the context of the broad range of AGN variability and of TDEs and other transient objects. In Figure~\ref{fig:DIS_comp_spec}, we show the optical spectrum of J234402 taken 35 days after the optical peak along with representative spectra from several other classes of objects, namely, changing-look AGN\footnote{Spectrum for J111536 made publicly available by \citet{YAN_2019}}, TDEs\footnote{\label{tns}Spectra available on the TNS server}, and a Hydrogen-rich SN\footref{tns}. Based on the strong X-ray emission, a supernova appears to be an unlikely explanation; however, the optical spectrum alone does not provide sufficient information to distinguish between the AGN and the TDE scenario.
Distinguishing between the two scenarios can be difficult as the observational characteristics can be ambiguous \citep[e.g.][]{DRAKE_2011,MERLONI_2015,NEUSTADT_2020,HINKLE_202108}. 

\subsection{J234402 and AGN ignitions}
\label{sec:discussion_agn}
\subsubsection{Physical parameters}
Under the assumption that the outburst in J234402 is caused by variability in an AGN, we can make estimates of some key properties of the system. To do so, we made use of relations between the broad line emitting region (BLR) and the continuum emission, which are well established for AGN. Based on our fit to the continuum and the established redshift of 0.10, we find a continuum luminosity at 5100 \AA\ of $\lambda$$L_\lambda$$\sim$$7.1$$\times$$10^{43}$ erg s$^{-1}$. Using the bolometric correction, log($L_\mathrm{bol})$=$4.89+0.91\times$log$(\lambda$L$_\lambda)$ \citep[cf.][]{RUNNOE_2012}, we find a value for the current isotropic bolometric luminosity of $L_\mathrm{bol}$$\sim$$6.2\times10^{44}$~erg~s$^{-1}$. We note that this estimate carries considerable uncertainty, as the bolometric corrections were derived for a broad population of AGN and do not necessarily translate to an extremely variable object such as J234402. 

We can compare the current $L_\mathrm{bol}$ with an estimate of the bolometric luminosity based on narrow the [OIII]$\lambda$5007 line. As the size of the narrow line region precludes any large variations in the timescale of the ignition event, $L_\mathrm{[OIII]}$ provides an indicator of the luminosity state $\sim$10$^3$ years prior to the ignition event. Using the bolometric correction of \citet{STERN_2012B}, we find $L_\mathrm{bol}^\mathrm{hist}$$\sim$$3.4\times10^{43}$~erg~s$^{-1}$. Based on narrow-line diagnostics (Figure~\ref{fig:OPT_bpt}), we associate the accretion phase that powered the [OIII] emission with AGN activity. As $L_\mathrm{bol}$$\sim$$20*$$L_\mathrm{bol}^\mathrm{hist}$, the nucleus is currently significantly brighter than during this historic AGN phase.

We can further compare the results of our optical fitting with values typical of AGN. The power-law continuum that is visible in the optical spectrum is somewhat steeper than the theoretical prediction for a thin accretion disk (with slope $\alpha$=$-2.52\pm0.01$). The theoretically predicted slope, $\alpha$=$-$2.33, has been found in difference spectra \citep[][]{KOKUBO_2014,MACLEOD_2016} of variable AGN, although averaged AGN spectra show a shallower slope of -1.54 \citep[][]{VANDENBERK_2001}. Based on the $L_{5100}$ to $L_{{\rm H}\alpha}$ scaling relationship established in \citet{GREENE_2005}, our measured value of $L_{5100}$ would correspond to $L_{{\rm H}\alpha}$=$3.5\times 10^{42}$ erg s$^{-1}$. This is significantly higher than the line luminosities we find in J234402 for which $L_{{\rm H}\alpha}$=$3.7\times 10^{41}$ erg s$^{-1}$, for the combined very-broad and broad-line components. The discrepancy suggests that the BLR is not as fully formed as it is in stably accreting AGN.

The tight scaling relationship between $\lambda$$L_{5100}$ and the light-travel distance to the line-forming region in AGN \citep[R$_{\mathrm{BLR}}$; e.g.][]{PETERSON_2004,BENTZ_2009,BENTZ_2013} allows us to make an estimate of R$_{\mathrm{BLR}}$. Using the scaling parameters established by \citet{BENTZ_2009} for H$\beta$ (their K=1.554 \& $\alpha$=0.546), we find R$_{\mathrm{BLR},H\beta}$=29.7 light days. Combining this radius with FWHM(H$\beta$), we are able to estimate a virial mass for the central black hole. For the combination of the broad and very broad Gaussian components, the FWHM(H$\beta$)=3321 km~s$^{-1}$. Adopting a virial factor $f$=1 \citep[cf.][]{PETERSON_2004} we find M$_\mathrm{BH}$$\sim$$10^{7.8}$ M$_{\odot}$. Using the scaling relation established by \citet{VESTERGAARD_2006}, we find a similar M$_\mathrm{BH}$$\sim$$10^{7.9}$~M$_{\odot}$. The associated Eddington luminosity, $L_{\mathrm{Edd}}$, for this mass is $8.0\times10^{45}$~erg~s$^{-1}$. Finally, combining the various derived properties, we find an Eddington ratio of $\lambda_{\mathrm{Edd}}$$\equiv$$L_\mathrm{bol}$/$L_{\mathrm{Edd}}$$\sim$$0.08$. As the bolometric correction assumes isotropic emission, a condition likely not fulfilled in the case of strong outflows, this value could represent a lower limit for $\lambda_\mathrm{Edd}$. We note that this method is based on the assumption of a virialised BLR, a condition possibly not met in J234402. Therefore, we emphasise that this mass estimate needs to be interpreted with caution. This is also evident in the choice of virial factor $f$. $f$ is dependent on both the orientation and the physical configuration of the BLR, which are unknown for J234402. Using the value $f$=4.3 \citep[cf.][]{GRIER_2013}, we find M$_\mathrm{BH}$$\sim$$10^{8.4}$~M$_{\odot}$, a significant difference from the previous estimate.

The mass estimate calibrated to an AGN sample can be compared with mass estimates based on different methods, to provide additional constraints. The black-hole mass in galaxies is known to correlate with the infrared luminosity of the galactic bulge \citep[][]{MARCONI_2003}. We make use of the updated correlation between M$_\mathrm{BH}$ and K-band magnitude provided in \citet{GRAHAM_2007}: log$\left( \mathrm{M}_\mathrm{BH}/\mathrm{M}_\odot\right) = -0.37(M_\mathrm{K}+24)+8.29$. In the 2MASS catalogue, the northern and southern sources are detected as separate objects, and we find $m_\mathrm{K}$=14.38 for J234402 (observed in 1998). We assume that at the time of measurement, the AGN contribution to the K-band flux is negligible. As the galaxy is not resolved, the 2MASS value represents an upper limit on the bulge luminosity and therefore on M$_\mathrm{BH}$. Using the 2MASS value, we find M$_\mathrm{BH}< 10^{8.3}$~M$_{\odot}$. In a different approach, we use the width of [OIII]$\lambda$5007 as an estimate of the bulge stellar-velocity dispersion $\sigma_*$, to apply the established M$_\mathrm{BH}$-$\sigma_*$ relation \citep[following e.g.][]{NELSON_2000}. This method assumes that the physical extent of the narrow-line-forming region is such that the velocity dispersion of the [OIII]-emitting gas is set by the gravitational potential of the bulge, making $\sigma_\mathrm{[OIII]}$ a good approximation of $\sigma_*$ \citep[cf.,][in galaxies with low M$_\mathrm{BH}$]{XIAO_2011}. Based on the Baade observation, we find $\sigma_\mathrm{[OIII]}$=169 km s$^{-1}$. In optical spectra, line broadening will always be a combination of instrumental and intrinsic broadening ($\sigma_{\rm tot}^2 = \sigma_{\rm J2344}^2 + \sigma_{\rm instr}^2$). To account for the strength of the instrumental effect for the [OIII]$\lambda$5007 line, we measured the instrumental broadening using the arc-lamp frame of the Baade observation. We find that the instrumental contribution to the line width is $\sim$1\%, whcih will be of negligible influence on the estimates black-hole mass. Using the scaling relation from \citet{TUNDO_2007}, we find M$_\mathrm{BH}$$\sim$$10^{7.9}$~M$_{\odot}$. The three methods of estimating the mass of the SMBH therefore provide consistent results, if we choose the lower value for the virial factor.

\subsubsection{Extreme X-ray variability in AGN}
Large amplitude X-ray brightening has been observed in the AGN NGC 2617 \citep{SHAPPEE_2014}, XMMSL1 J061927.1-655311 \citep[][]{SAXTON_2014}, HE 1136--2304 \citet{PARKER_2016}, iPTF 16bco \citep{GEZARI_2017}, NGC 1566 \citep[][]{PARKER_2019,OKNYANSKY_2020}, 1ES 1927+654 \citep{TRAKHTENBROT_2019B,RICCI_2020,RICCI_2021}, SDSS J155258+273728 \citep[][]{AI_2020}, and AT2019pev \citep[][]{YU_2022}. In all of these objects, the X-ray brightening was accompanied by rapid optical and UV flux increases (where UV observations were available). In most objects, the X-ray brightening was accompanied by the appearance of broad Balmer emission lines in the optical spectrum. However, broad emission lines in HE 1136--2304 had \emph{disappeared} post-brightening, compared to an archival spectrum taken 11 years earlier. The order of magnitude increase in 0.3--10 keV flux in UGC 2332 \citep[][]{WANG_2020} was matched by a change in optical spectrum from type 2 to type 1.8 (i.e. appearance of broad H$\alpha$), reversing a transition from 1.5 to 2 in the previous decades. 

The persistence of the new, brighter state and any emission lines differs from object to object --- in NGC 2617, the persistence of UV emission years later suggests a long-lasting change of accretion state \citep[][]{OKNYANSKY_2017}, whereas in NGC 1566 and 1ES 1927+654, the flux faded over months--years. When sufficiently sampled, the shapes of the light curves therefore provide some distinguishing power between AGN and TDEs. However, the most important distinguishing characteristic for all these objects is the hardness of the X-ray spectrum. In all events discussed above, the X-ray spectrum had a power-law component with $\Gamma$$\lesssim$2, thus firmly identifying an AGN contribution to the emission. \citet[][]{NODA_2016} found that in the Seyfert 1.5 NGC 3516, there is in fact a positive correlation between B-band flaring and the strength of the $\Gamma$$\sim$1.7 power-law component in the 2--45 keV spectrum. The very soft spectrum of J234402 therefore stands out.

\subsubsection{Soft X-ray ignitions in AGN}
\label{sec:discussion_agn_soft}
The nearby galaxy NGC 3599 \citep[][]{ESQUEJ_2007} brightened in X-rays by a factor of 150 between 1993 and 2003 and showed a particularly soft spectrum ($\Gamma$$\sim$3 for a fit with a single power law). The bright state lasted at least 18 months and was followed by a decline lasting several years. Based on the timescales involved, \citet[][]{SAXTON_2014} concluded a fast-rising TDE is unlikely, making AGN variability based on a disk instability (see below) the more likely cause. IC 3599 \citep[][]{BRANDT_1995,KOMOSSA_1999} has shown X-ray outbursts in 1990 and 2010, which have been interpreted both as repeated partial tidal stripping of a star \citep[][]{CAMPANA_2015} and intrinsic AGN variability \citep[][]{GRUPE_2015}. The ROSAT spectrum of the first outburst was notably soft ($\Gamma$$\sim$4.8), although the \textit{Swift-XRT} spectra of the second outburst indicate that a fit with a power law+blackbody (with power-law index $\Gamma$$\sim$2.6) is perhaps more suitable. This harder-when-fainter behaviour, typical of AGN, is not observed in J234402, although our continuing observations will allow us to confirm this.

These soft X-ray ignitions differ from J234402 in several key aspects: (1) they are shorter lived, (2) they are associated with smaller SMBHs (typically M$_\mathrm{BH}$$\sim$$10^{5-6}$ M$_\odot$ with the exception of NGC 3599, which has M$_\mathrm{BH}$$\sim$$10^{8}$ M$_\odot$ and an evolution over months--years), and (3) all objects for which optical spectroscopy is  available lack strong broad emission lines in their optical spectra.

\subsubsection{Timescales}
\label{discussion_timescales}
To qualify the nature of the accretion process, the timescales of the changes provide some of the most helpful constraints. Although one can define many model-dependent timescales for accretion flows, they generally fall into a few broad categories: the light crossing timescale ($t_{lc}$), the dynamical or orbital timescale ($t_{dyn}$), the thermal timescale ($t_{th}$) related to cooling or heating of the disk, and the viscous timescale ($t_{\nu}$) at which large changes can propagate through the disk through viscosity. For the standard thin-disk model and using the parametrisation provided in \citet[][]{STERN_2018}, we can determine the aforementioned timescales in J234402. We use $M_\mathrm{BH} = 10^{7.9}$ M$_\odot$, set the viscosity parameter to $\alpha = 0.03$ and the ratio between the disk scale height and the radius ($H/R$) to 0.05 (these values for $\alpha$ and $H/R$ are typical for a geometrically thin disk). As the best-constrained timescales for J234402 are based on optical data, we evaluate the theoretical timescales at 100 $r_g$ --- this represents the typical size of the inner optically emitting region\footnote{depending on the contribution of reprocessed X-ray and UV radiation to the optical output \citep[see e.g.][for a discussion on this topic]{NODA_2016}.}. With these considerations, we find the following estimates for the timescales relevant for accretion flows:
\begin{subequations}\label{eq:timescales}
\begin{align}
    t_{lc} & = \frac{R}{c} \sim 10.9\hspace{.1cm}\mathrm{hours} \\
    t_{dyn} & \sim \sqrt{\frac{R^3}{GM}} \sim 4.3\hspace{.1cm}\mathrm{days} \\
    t_{th} & \sim \frac{1}{\alpha}*t_{dyn} \sim 144\hspace{.1cm}\mathrm{days} \\
    t_{\nu} & \sim \left(\frac{H}{R}\right)^{-2}*t_{th} \sim 158\hspace{.1cm}\mathrm{years}.
\end{align}
\end{subequations}
We can compare these values with the observed evolution in J234402. The rapid rise time (Phase I) is $\sim$30 days and the total time to peak (Phase I+II) is $\sim$60 days. The decay timescale is of course not well constrained by the current dataset (this will be covered in our Paper~II); however, we can use $t_{sc}$ (Equation~\ref{eq:decay} in Section~\ref{sec:mw_opt_lc}) as a first estimate. $t_{sc}$ is $\sim$80 days for the ATLAS $c$-band data and $\sim$186 days for the $o$-band data. These values best match our estimates for the timescale $t_{th}$, although for the rise time the dynamical timescale is also a good match. 

The defining timescale for the standard thin accretion disk is the viscous timescale; however, it has been found that many forms of AGN variability operate on timescales too small to match this description \citep[see e.g.][]{LAWRENCE_2018}. Several instabilities that operate on $t_{th}$ have been proposed to explain extreme changes in AGN accretion rates \citep[see e.g. the discussion in][]{STERN_2018}, which could reasonably explain the outburst in J234402.

One possible mechanism for speeding up large changes is the Lightman-Eardley (LE) disk instability \citep[][]{LIGHTMAN_1974} in combination with a truncated accretion disk. In this scenario, the disk is truncated at a given inner radius, resulting in an inner region that slowly ($\sim$$t_\nu$) fills with material until the LE instability triggers a rapid ($\sim$$t_{th}$) ignition and heating of the disk that produces a soft, thermal X-ray flare. The inner region is then drained once more as the inner disk is accreted, resulting in the decay phase in the light curve. \citet[][]{SAXTON_2015} find that such a process is a possible explanation for the variability observed in NGC~3599. The decay timescale in this scenario is constrained to be greater than $t_{dyn}$, but also to be smaller than the rise timescale. The first condition is easily matched by J234402; however, the second condition does not appear to hold. Although the optical brightening in J234402 matches a thermal timescale for the system, the relatively slow decay implies that an LE instability cannot fully explain the observed behaviour. 

\subsection{J234402 and TDEs}\label{sec:discussion_tde}
\subsubsection{Comparing physical parameters}
Following the criteria on X-ray behaviour set out in \citet[][]{AUCHETTL_2018}, J234402 is a good TDE candidate, as the spectrum is very soft and shows limited spectral evolution in the first months following the flare. A soft X-ray spectrum best fit with a blackbody model is common in TDEs \citep[][]{GEZARI_2021}. The rest-frame luminosity of $L_\textrm{0.2--2 keV}$=$10^{44.9}$ erg s$^{-1}$, which is derived from the eRASS2 observation, is quite high for a non-jetted TDE (Section~\ref{sec:mw_fermi}). The luminosity function derived by \citet[][]{AUCHETTL_2018} sharply cuts off around $10^{44}$ erg s$^{-1}$. The estimated SMBH mass of $10^{7.9}$ M$_\odot$ is quite near the Hills mass \citep[][]{HILLS_1975} of a non-spinning BH with $M_{\mathrm{BH}}= 10^{8}$ M$_\odot$, which is the theoretical limit for a tidal disruption of a star outside the black hole's event horizon. However, black-hole spin could allow for black holes with masses greater than the Hills mass to produce very luminous TDEs \citep[][]{LELOUDAS_2016,MUMMERY_2020}. The X-ray light curve (Figure~\ref{fig:xray_combined}) shows a clear decay, albeit not strictly monotonic. Short-term X-ray variability in TDEs is a known phenomenon \citep[e.g.][]{WEVERS_2019,VANVELZEN_2021}. J234402 would therefore be an unusually bright but not an unusually variable TDE candidate.

The ratio of the [OIII]$\lambda$5007 to 0.2--2 keV X-ray luminosity for J234402 is approximately $10^{4.1}$. This is significantly larger that the $L_{\rm 0.3-20 keV}$/$L_{\rm [OIII]}$ ratio of $10^{2.0}$ found for local type 1 AGN \citep[][]{HECKMAN_2005}, but is consistent with the values found for TDE candidates \citep[][]{SAZONOV_2021}. This result is in agreement with the large discrepancy that we find between $L_\mathrm{bol}^\mathrm{hist}$ derived from [OIII] and the current $L_\mathrm{bol}$ (based on the optical spectrum, see Section~\ref{sec:discussion_agn}): the current luminosity significantly surpasses the luminosity associated with any previous AGN activity. 

\begin{figure}
    \centering
    \includegraphics[trim=0.5cm 0.5cm 1.5cm 1.7cm,clip,width=\linewidth]{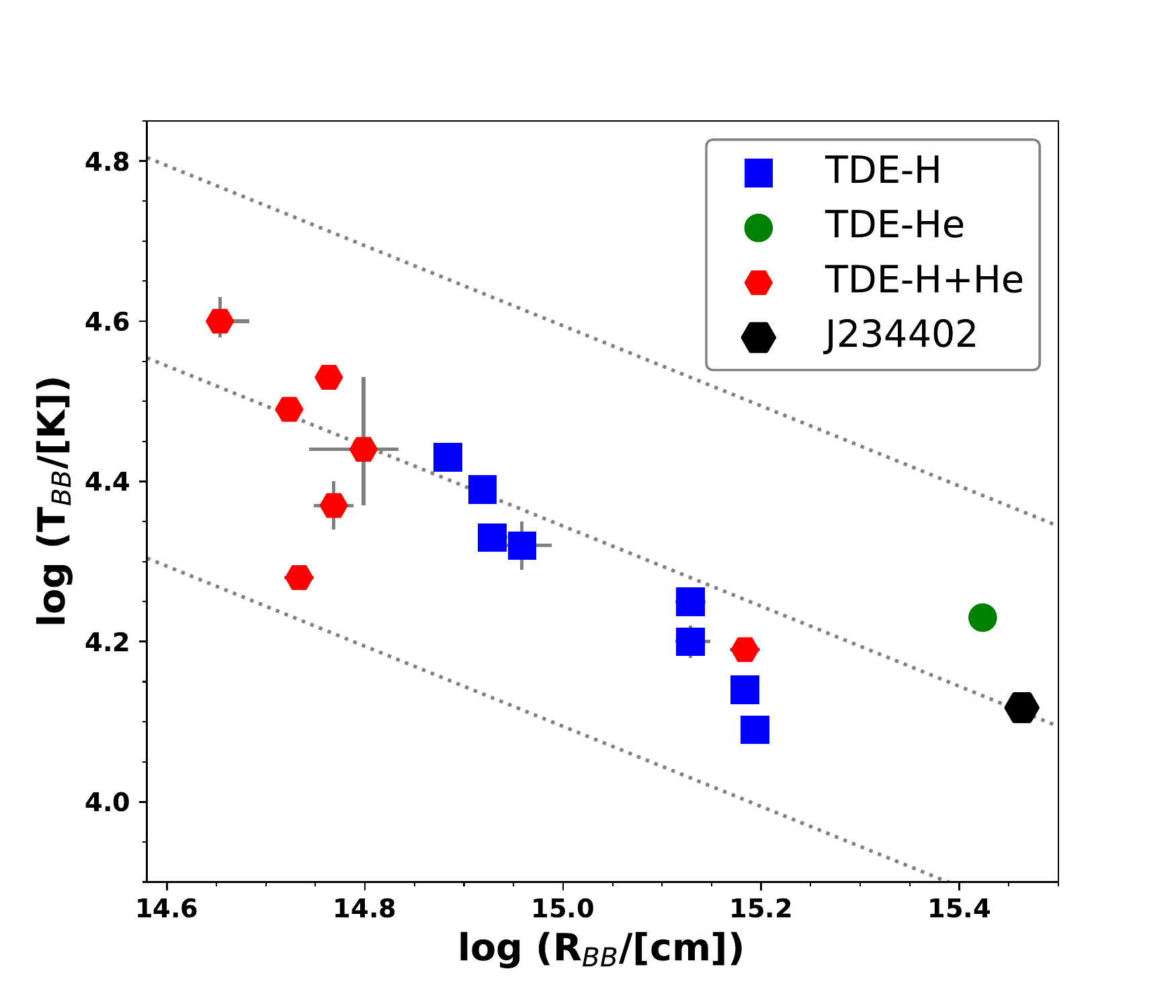}
    \caption{Derived blackbody temperatures and blackbody radii of several TDEs compared to those of J234402. TDE-H and THD-He refer to TDEs with only Hydrogen or only Helium lines in their spectrum, respectively (see Section~\ref{sec:discussion_tde} for further detail). The dashed grey lines show the expected log T$_{BB} \propto -\frac{1}{2}$log R$_{BB}$ behaviour for several example values of L$_\textrm{bol}$. J234402 lies at the relatively cool and large ends of the distribution. The TDE values were all derived for the period around $t_{peak}$, whereas the values for J234402 were derived from spectra taken over a period approximately 25 days after the peak. The TDE data were published in \citet{VANVELZEN_2021}.}
    \label{fig:DIS_tbb_rbb}
\end{figure}

Under the assumption that the optical continuum flux is driven by blackbody emission, we estimate the blackbody temperature (T$_\mathrm{BB}$) and bolometric luminosity from the optical spectra (see Figure~\ref{fig:opt_specfit_bb}) and derive the blackbody radius (R$_\mathrm{BB}$). Using the average values from the four spectra, we find log(T$_\mathrm{BB}$/[K])=4.1 and log(R$_\mathrm{BB}$/[cm])=15.5. Figure~\ref{fig:DIS_tbb_rbb} shows a comparison of our values with those derived for the TDE sample presented in \citet[][]{VANVELZEN_2021}. J234402 appears on the larger and cooler ends of the ranges of values. However, we note that the values for J234402 are based on observations taken $\sim$25 days after the peak, whereas the other measurements are around $t_{peak}$. J234402 may have had more time to expand and cool. Interpreting the optical emission from J234402 as a blackbody therefore appears consistent with the results from other TDEs.  We find $L_\mathrm{bol,BB}$=$1.7\times 10^{44}$ erg s$^{-1}$, in approximate agreement with the result from Section~\ref{sec:discussion_agn}. To calculate the ratio between the luminosity based on the optical data, $L_\mathrm{bol,BB}$, and the soft X-ray luminosity, we make use of the \textit{XMM-Newton} data as the observation periods nearly coincide. We find  $L_\mathrm{bol,BB}/L_X$$\approx$0.7, which is on the low end for most TDEs \citep[compared with Figure 10 in][]{VANVELZEN_2021}.

\subsubsection{Optical evolution}
The optical identification of TDEs has mostly been based on large photometric surveys \citep[][refs]{VANVELZEN_2021,GEZARI_2021}. The timescale of the rapid rise in J234402 (Phase I, before the plateau) of $\sim$30 days fits within the observed range of rise times for TDEs. We find that the rise time in the sample of \citet[][]{VANVELZEN_2021} varies from $\sim$2 to 32 days. J234402 is therefore at the long end of this range. The temporary halt in the flux increase (Phase II) is unusual in the light curves of TDEs, lacking a clear equivalent in observed objects. The optical decline exhibits a power-law-like slope that is consistent with TDEs \citep[][]{GEZARI_2021}. 

The spectroscopic presentation of optically selected TDEs shows a broad distribution in the strength of H and He lines. Generally, TDEs can be classified into objects that develop broad H$\alpha$ and H$\beta$ lines (TDE-H), those that develop Balmer lines as well as broad He II $\lambda$4686 (TDE-H+He), and those that only show broad He features \citep[TDE-He;][]{ARCAVI_2014,LELOUDAS_2016,VANVELZEN_2021}. Using this classification system, J234402 would fall into the TDE-H category. Bowen fluorescence, which has been associated with the TDE AT2019qiz \citep[][]{NICHOLL_2020} and the nuclear transient AT2017bgt \citep[][]{TRAKHTENBROT_2019B}, is not apparent in J234402 (N~III~$\lambda$4640 is in range for our spectroscopy). 

One of the most distinctive features of J234402's spectrum is the strongly asymmetric line profile of the Balmer lines. The FWHM of the broadest components is $\sim$$10^3$ km s$^{-1}$, which is of the same order of magnitude as the transient and blueshifted broad emission features detected in the TDEs AT2019qiz \citep[H$\alpha$;][]{NICHOLL_2020} and PS1-10jh \citep[He~II~$\lambda$4686;][]{GEZARI_2015}. In both cases, the offset of the broad components is associated with an outflow, even matching outflow signatures in radio observations in AT2019qiz. 

Coronal emission lines, such as [ArXIV]$\lambda$4414, [FeIX]$\lambda$5304, [FeVII]$\lambda$6088, have been associated with TDEs \citep[][]{KOMOSSA_2008,WANG_2012} as well as variable AGN \citep[][]{BRANDT_1995,GRUPE_1995}. In a subset of extreme coronal line emitters, the coronal emission has been accompanied by the presence of strong Balmer emission lines \citep[][]{WANG_2012}. Given the strength of the X-ray emission is J234402, we investigated the possibility of a similar combination of these emission features in our optical spectra. We find no evidence of coronal line emission. The lack of any coronal lines, or indeed of lower-ionisation FeII emission features, could indicate that the available gas is relatively metal-poor. Alternatively, the lines will take more time to develop.

\subsection{IR evolution}
The observed increase in the IR light curve, associated with the X-ray and optical outburst (Section~\ref{sec:mw_phot_wise}), provides additional information about the circumnuclear environment. One possibility is that the IR emission is part of the same outburst as the optical flare. As we estimate the mass of SMBH to be rather high, it is feasible that the outer reaches of the accretion disk have a sufficiently low effective temperature to emit significantly in the IR. In this case the increased emission is the result of a heating up of an existing accretion disk, as the effect of the accretion-rate increase in the inner disk travels outwards. A different interpretation is that the IR emission is the result of reprocessing of UV and optical emission by dust, which is commonly seen in AGN  \citep[for a review, see e.g.][] {NETZER_2015} and has also been observed in a small number of TDEs \citep[][]{VANVELZEN_201609,JIANG_2017,JIANG_2021}. Under this assumption, we can estimate a minimum response timescale for the circumnuclear dust to the outburst. The interval between $t_{start}$, the beginning of the optical ignition event, and the most recent WISE epoch is approximately 68 days. If we further assume that the reprocessing of incoming radiation by the dust is immediate, we can relate this maximum response time to a light-travel distance from the central engine of $1.8\times10^{17}$ cm (0.06 pc). We can compare this distance to the mean sublimation radius \citep[cf.][]{BARVAINIS_1987,NETZER_2015} within which the incident UV flux is strong enough to destroy dust grains by photodissociation:
\begin{equation}\label{eq:dust_subl}
    R_{subl} \approx 0.5 \textrm{ pc} \hspace{2mm} \Bigg(\frac{L_\textrm{bol}}{10^{45}\textrm{erg s}^{-1}}\Bigg)^{1/2} \hspace{2mm} \Bigg(\frac{\textrm{T}_{sub,X}}{1500 \textrm{K}}\bigg)^{-2.6} \hspace{2mm} f(\theta),
\end{equation}
where T$_{sub,X}$ is the sublimation temperature for different types of dust grains, typically 1800~K for carbonaceous grains and 1500~K for silicate grains \citep[][]{NETZER_2015}, and the angular term $f(\theta)$ defines the anisotropy of the nuclear emission. Using our estimate for $L_\textrm{bol}$ of $6.2\times10^{44}$ erg~s$^{-1}$, T$_{sub,Si}$, and $f(\theta)$=1, we find a sublimation radius of $1.2\times10^{18}$ cm (0.39 pc), much larger than the light-travel distance. A refinement of this approach would be to take into account the physical distribution of the dust: if the observed response originates in dust that is located along our line of sight to the nucleus, the light travel time could be longer than in the rough estimate given above. In fact, the profile of the IR light curve in the months--years following the outburst can help track the distribution of the dust, based on varying response times. We will investigate this evolution in follow-up work.

In a sample of 23 optically selected TDEs, \citet[][]{JIANG_2021} found that most TDEs showed no response in the IR emission, except for those that appeared to occur within AGN. A possible interpretation for this correlation is that the covering factor of dusty regions in inactive galaxies is too small for any detectable reprocessing by dust to occur. A significant accretion rate may be necessary to support a thick, dusty structure \citep[][]{VANVELZEN_2016}. IR echos from TDEs in AGN have also been observed in radio-selected TDEs \citep[e.g.][]{MATTILA_2018}. The W1-W2 colour in J234402 around the time of peak is 0.26, which can be associated with a blackbody temperature $\sim$2600~K. During the last observation before the ignition event, W1-W2=0.35, which corresponds to a temperature of $\sim$2100~K. It therefore appears that the brightening is associated with an increase in the dust temperature. A change in dust temperature has been associated with TDEs showing IR dust echos \citep[][]{VANVELZEN_2011,JIANG_2017}, whereas three changing-look AGN that were selected based on changes in their WISE magnitudes \citep[][]{STERN_2018,ROSS_2018} showed no significant temperature changes, on a timescale of years. 

\subsubsection{A TDE in an AGN}
\label{sec:discussion_tde_in_agn}
Several characteristics of J234402 best match a TDE, in particular the soft X-ray spectrum and the onset of a rapid decay. However, the optical spectrum indicates the presence of an AGN. In recent years, more objects with a mix of AGN and TDE qualities have been observed, presenting a challenge to classification. AT2017bgt \citep[][]{TRAKHTENBROT_2019A} and ASASSN-18jd \citep[][]{NEUSTADT_2020} both exhibited strong increases in the optical, hard X-ray spectra, and combinations of spectral features observed in AGN (narrow [OIII]) and in TDEs (e.g. Bowen fluorescence of NIII$\lambda$4640). The longevity of the new spectral features ($>$1 year) indicates these may be a new type of transient \citep[][]{TRAKHTENBROT_2019A}. The transient ASASSN-17jz \citep[][]{HOLOIEN_2022} was classified as a possible SN~IIn outburst that may have occurred in or near the accretion disk of an existing, low-luminosity AGN and may have triggered an increase in the accretion rate. Like J234402, the object has a soft X-ray spectrum ($\Gamma$$\sim$3.4), strong Balmer lines in its optical spectra, and no evidence of Bowen fluorescence. A very broad ($\sim$6000 km s$^{-1}$) H$\beta$ component faded on a timescale of years. We inspected the NEOWISE-R observations for ASASSN-17jz and found that a sharp rise and subsequent decay are clearly visible in the IR light curve, coincident with the ignition event. This could present an additional similarity to J234402 and is consistent with the notion that a pre-existing disk is a requirement for a strong IR echo. 

There are several reports of possible TDEs occurring within objects classified as AGN, including ASASSN-14li \citep{VANVELZEN_2016}, PS16dtm \citep[][]{BLANCHARD_2017}, 1ES~1927+654 \citep[][]{TRAKHTENBROT_2019B,RICCI_2020,RICCI_2021}, and SDSS J022700.77-042020.6 \citep[][]{LIU_2020}. The behaviours of these objects are quite diverse. The optical light curves of 1ES~1927+654 and SDSS J0227-0420 are similar to that of J234402; however, PS16dtm reaches a plateau in the light curve that lasts several months. Both PS16dtm and 1ES~1927+654 dimmed in X-rays following the optical ignition, due to the disappearance of the power-law component in the spectrum, though in the case of 1ES~1927+654, the X-rays and the power-law component recover on a timescale of months. The decay of the X-ray light curve in ASASSN-14li is also slower than observed in J234402. The optical spectra of 1ES~1927+654 show a combination of narrow [OIII] lines and strong broad Balmer lines. The Balmer lines in 1ES~1927+654 appeared on a timescale of months (this is not constrained for J234402, but the presence of strong Balmer lines $\sim$70 days after $t_{peak}$ matches this timeline). In contrast, PS16dtm shows little spectral evolution, except the development of a complex of FeII lines, which is not observed in J234402. A possible explanation for this difference is that PS16dtm was classified as a Narrow-Line Sy1, characterised by a low M$_\textrm{BH}$ and strong Iron emission, in contrast to the high M$_\textrm{BH}$ of J234402. We therefore do not find a clear equivalent to J234402 among other ambiguous transients, although ASASSN-17jz shares some significant characteristics.

Summarising, we can say that the soft X-ray spectrum and rapid onset of the decay observed at all wavelengths indicate that J234402 was likely powered by a TDE. The relative strength of the high-ionisation [OIII] lines, compared to the [NII] and [SII] lines, in the optical spectra in J234402 indicate the (historical) presence of an AGN. The non-detection in ROSAT and the lack of any clear power-law component in our new X-ray spectra indicate that J234402's AGN phase was likely in the past. However, the possible low-level variability in the optical light curve, visible in the ATLAS data, means we cannot exclude the possibility that the SMBH in J234402 was still actively accreting at a low level in the years directly prior to the ignition. Together this paints a scenario in which a TDE occurred in an environment that was shaped by AGN activity. 

For TDEs that occur in AGN, the accretion process powering the emission can be significantly more complex than in `secular' TDEs, owing to the interaction between the stellar debris with the pre-existing disk \citep[][]{CHAN_2019}. The different kinematic components we detect in the broad Balmer lines are indicative of a rapidly evolving system, with the broadest (i.e., the innermost) components providing strong evidence of an outflow. We speculate that the apparent stratification in the BLR is affected by the AGN environment of the TDE --- fast-orbiting clouds close to the central engine are generated in the event itself, whereas the outer BLR could be caused by irradiation of a pre-existing structure.

\section{Conclusion and outlook}
\label{sec:conclusion}
We have presented the results of the eROSITA detection and subsequent multi-wavelength follow-up of a significant extragalactic X-ray ignition in J234402, in which the 0.2--2 keV flux increased by at least a factor of 150 in the six-month interval between eROSITA scans. The spectrum is very soft and best fit with a dual blackbody (T$\sim$10$^6$ K). The X-ray flux shows strong variability on timescales from hours to days. These short-term fluctuations occur within a trend of overall decline in the two months following the detection. Optical photometry shows that prior to the X-ray detection, the emission from J234402 showed a rapid rise of $\sim$3 mag on a timescale of weeks, which is followed by a plateau and decline in the following months. This decline is matched in the UV. Our follow-up optical spectroscopy shows that the source has a blue continuum and strong, broad Balmer emission lines, as well as narrow high-ionisation [OIII] lines. We do not observe any optical spectral evolution over the course of our follow-up. The IR emission shows an increase that is correlated with the optical flare.

Based on the optical spectra and using AGN scaling relations, we derive a bolometric luminosity of $6.1\times 10^{44}$ erg~s$^{-1}$ and a black hole mass of $10^{7.9}$ M$_\odot$. The ignition event has characteristics in agreement with both a TDE and an AGN; however, the soft X-ray spectrum and the onset of a rapid decline in the light curve lead us to identify J234402 as a likely TDE within a low-luminosity or turned-off AGN. The strength of the high-ionisation narrow lines, as evident in the BPT diagnostic, suggests that J234402 was in an AGN phase as recent as a few millennia ago. We have limited our discussion to the data before January 2021 when J234402 went into Sun block. Our data cover the rise and peak of the X-ray and optical light curves. To confirm our interpretation of the accretion event, we will make use of further follow-up data, which we are currently gathering. 

We expect that our continued observations will allow us to better constrain the interaction between the ignition event and the surrounding nuclear medium. Photometric monitoring will enable us to find the decay timescale and to search for any deviations from the monotonic decay in the UV and optical that is expected for TDEs. We expect the X-ray variability to diminish over time, as we associate it with the transient emission originating from the interaction of the accreting stellar debris and the disk. Using the decay light curve, we will also be able to put constraints on the properties of the stellar progenitor of the TDE. By tracking the evolution of the broad emission lines, we will be able to see if the distinct kinematic structures we identified will evolve differently, or if a shared origin in a single outflow is a more likely scenario. Finally, if the IR response is caused by a dust echo, we expect the emission to brighten further before also diminishing. The analysis of the combined data, including the period after Sun block and new radio observations, is underway and will be the topic of follow-up papers.

\begin{acknowledgement}
The authors thank the anonymous reviewer for their insightful and constructive suggestions.
DH acknowledges support from DLR grant FKZ 50 OR 2003. MK is supported by DFG grant KR 3338/4-1.
AM, TS, and SK acknowledge full or partial support from Polish
Narodowym Centrum Nauki grants 2016/23/B/ST9/03123, 2018/31/G/ST9/03224, and 2019/35/B/ST9/03944.
AG was funded by the German Science Foundation (DFG grant number KR 3338/4-1). MG is supported by the EU
Horizon 2020 research and innovation programme under grant agreement No 101004719.
The authors wish to express their thanks to J.~Wilms and A.~Schwope for their contributions to many useful discussions. 
This work is based on data from eROSITA, the soft X-ray instrument aboard SRG, a joint Russian-German science mission supported by the Russian Space Agency (Roskosmos), in the interests of the Russian Academy of Sciences represented by its Space Research Institute (IKI), and the Deutsches Zentrum für Luft- und Raumfahrt (DLR). The SRG spacecraft was built by Lavochkin Association (NPOL) and its subcontractors, and is operated by NPOL with support from the Max Planck Institute for Extraterrestrial Physics (MPE). The development and construction of the eROSITA X-ray instrument was led by MPE, with contributions from the Dr. Karl Remeis Observatory Bamberg \& ECAP (FAU Erlangen-Nuernberg), the University of Hamburg Observatory, the Leibniz Institute for Astrophysics Potsdam (AIP), and the Institute for Astronomy and Astrophysics of the University of Tübingen, with the support of DLR and the Max Planck Society. The Argelander Institute for Astronomy of the University of Bonn and the Ludwig Maximilians Universität Munich also participated in the science preparation for eROSITA.
This work was supported in part by NASA through the \textit{NICER} mission and the Astrophysics Explorers Program. \textit{NICER} data used in this work were gathered under a Guest Observer (GO) approved programme and \textit{NICER} DDT time augmented the GO-approved time significantly.
We acknowledge the use of public data from the Swift data archive (ObsIDs: 13946001, 13946002, 13946004-006).
This work is based on observations obtained with XMM-Newton (ObsID: 0862770101), an ESA science mission with instruments and contributions directly funded by ESA Member States and NASA.
The ROSAT Project was supported by the Bundesministerium f\"{u}r Bildung und Forschung (BMBH/DLR) and the Max-Planck-Gesellschaft.
This research has made use of the NASA/IPAC Infrared Science Archive, which is funded by the National Aeronautics and Space Administration and operated by the California Institute of Technology
The \textit{Fermi} LAT Collaboration acknowledges generous ongoing support from a number
of agencies and institutes that have supported both the development and the operation of
the LAT as well as scientific data analysis. These include the National Aeronautics and
Space Administration and the Department of Energy in the United States, the Commissariat
à l'Energie Atomique and the Centre National de la Recherche Scientifique / Institut
National de Physique Nucléaire et de Physique des Particules in France, the Agenzia
Spaziale Italiana and the Istituto Nazionale di Fisica Nucleare in Italy, the Ministry of
Education, Culture, Sports, Science and Technology (MEXT), High Energy Accelerator
Research Organization (KEK) and Japan Aerospace Exploration Agency (JAXA) in Japan,
and the K. A. Wallenberg Foundation, the Swedish Research Council and the Swedish
National Space Board in Sweden.
Additional support for science analysis during the operations phase is gratefully
acknowledged from the Istituto Nazionale di Astrofisica in Italy and the Centre National
d'Etudes Spatiales in France. This work performed in part under DOE Contract DE-AC02-
76SF00515.
This work has made use of data from the Asteroid Terrestrial-impact Last Alert System (ATLAS) project. The Asteroid Terrestrial-impact Last Alert System (ATLAS) project is primarily funded to search for near earth asteroids through NASA grants NN12AR55G, 80NSSC18K0284, and 80NSSC18K1575; byproducts of the NEO search include images and catalogs from the survey area. This work was partially funded by Kepler/K2 grant J1944/80NSSC19K0112 and HST GO-15889, and STFC grants ST/T000198/1 and ST/S006109/1. The ATLAS science products have been made possible through the contributions of the University of Hawaii Institute for Astronomy, the Queen’s University Belfast, the Space Telescope Science Institute, the South African Astronomical Observatory, and The Millennium Institute of Astrophysics (MAS), Chile.
We acknowledge ESA Gaia, DPAC and the Photometric Science Alerts Team (http://gsaweb.ast.cam.ac.uk/alerts).
This paper includes data gathered with the 6.5 meter Magellan Telescopes located at Las Campanas Observatory, Chile.
Parts of this work are based on observations made with ESO telescopes at La Silla Paranal Observatory under ESO programme 105.20UT.001.
Some of the observations reported in this paper were obtained with the Southern African Large Telescope (SALT) under program 2020-2-MLT-008 (PI: A. Markowitz). Polish participation in SALT is funded by grant No. MNiSW DIR/WK/2016/07.
This publication makes use of data products from the Two Micron All Sky Survey, which is a joint project of the University of Massachusetts and the Infrared Processing and Analysis Center/California Institute of Technology, funded by the National Aeronautics and Space Administration and the National Science Foundation.
The Digitized Sky Surveys were produced at the Space Telescope Science Institute under U.S. Government grant NAG W-2166. The images of these surveys are based on photographic data obtained using the Oschin Schmidt Telescope on Palomar Mountain and the UK Schmidt Telescope. The plates were processed into the present compressed digital form with the permission of these institutions.
The UK Schmidt Telescope was operated by the Royal Observatory Edinburgh, with funding from the UK Science and Engineering Research Council (later the UK Particle Physics and Astronomy Research Council), until 1988 June, and thereafter by the Anglo-Australian Observatory. The blue plates of the southern Sky Atlas and its Equatorial Extension (together known as the SERC-J), as well as the Equatorial Red (ER), and the Second Epoch [red] Survey (SES) were all taken with the UK Schmidt.
\end{acknowledgement}

\bibliographystyle{aa} 
\bibliography{references.bib}

\begin{appendix}
\section{X-ray observations}
\label{sec:app_xray}
\subsection{\textit{XMM-Newton} analysis}
\begin{figure*}[!h]
    \centering
    \includegraphics[width=0.55\linewidth]{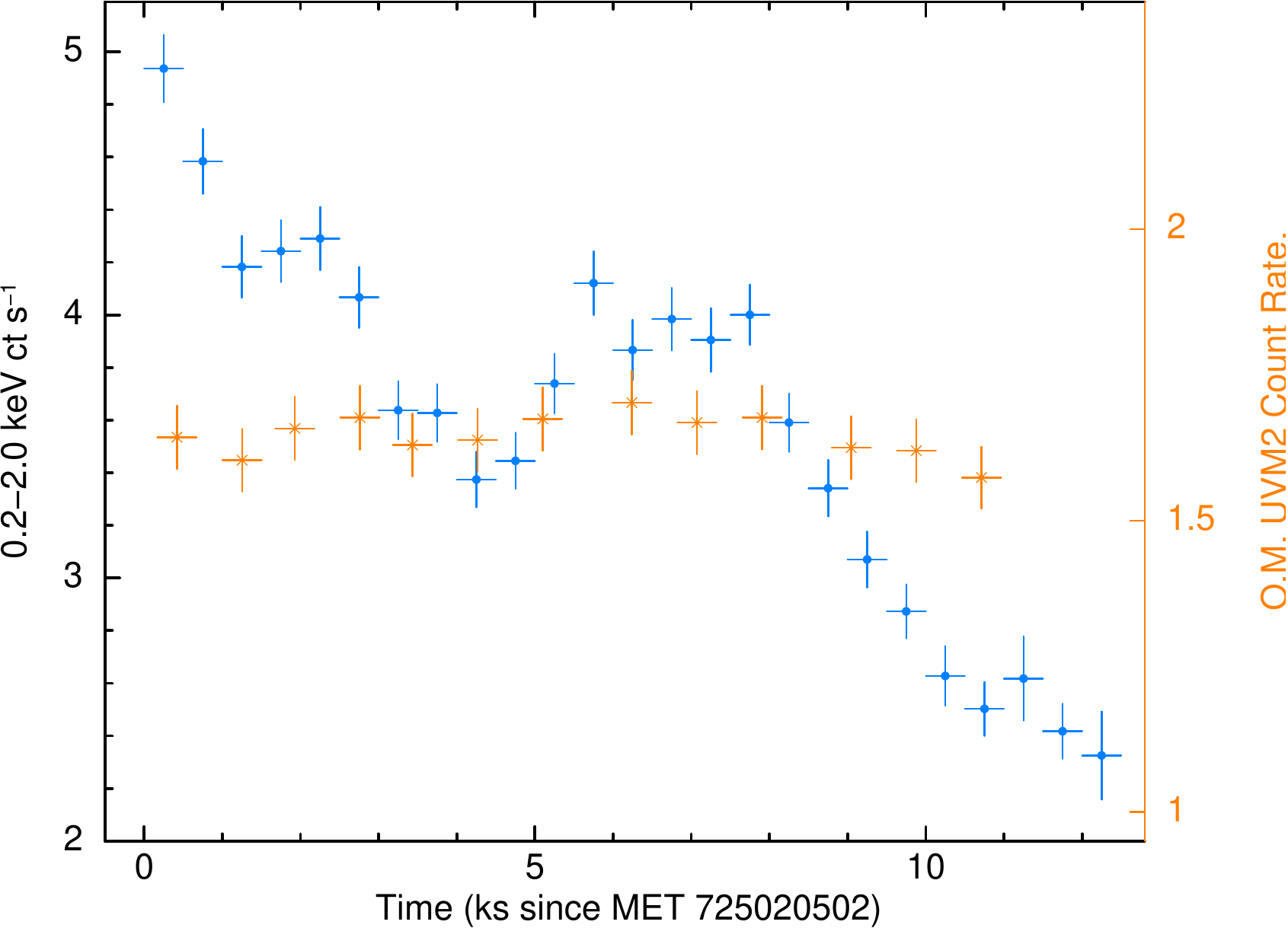}
    \caption{\textit{XMM-Newton} count rate light curves for EPIC pn 0.2--2.0 keV (500 s bins; \textit{blue}) and for Optical Monitor UVM2 fast mode (800~s bins; \textit{orange}). MET refers to Mission Elapsed Time, and the MET zeropoint used for the x-axis corresponds to 10:33:53 on 23 December 2020. Both datasets are plotted in counts per second. Whereas the X-ray emission is clearly strongly variable over the duration of the observation, the optical light curve shows no significant variability on this timescale.}
    \label{fig:xmm_lc}
\end{figure*}

Figure~\ref{fig:xmm_lc} shows the combined X-ray and UV counts data of the \textit{XMM-Newton} observation. We test a range of spectral models in  \textsc{Xspec} (version 12.11.1d) using $\chi^2$ statistics. In parallel, we test the same models using the nested sampling algorithm \textsc{multinest} v.3.10 \citep{SKILLING_2004,FEROZ_2009} via the Bayesian X-ray Analysis and PyMultinest packages for \textsc{Xspec} \citep[BXA, version 3.31, using Xspec version 12.10.1;][see section~\ref{sec:xray_xmm}]{BUCHNER_2014}
Both methods yield highly similar results in terms of model goodness of fit. In addition we find that for the majority of model parameters, best-fit values and 90$\%$ confidence level uncertainties obtained using both methods are consistent. We therefore included only the Bayesian evidence in the discussion in Section~\ref{sec:xray_xmm}; however, we present a full overview of our fitting results in Table~\ref{tab:xray_xmmfit}.

For the \textit{XMM-Newton} data we use $\Delta\chi^2=\chi^2_{\rm min}+2.71$ for the least-squares fitting, which corresponds to a 90$\%$ confidence level for one parameter when errors are symmetric. We use the 5th and 95th percentiles of the posterior distribution for Bayesian fits.

\begin{table}[!h]
\renewcommand*{\arraystretch}{1.1}
    \centering
    \caption{
    Overview of the best-fitting models to the \textit{XMM-Newton} data.
    }
    \begin{tabular}{lccl}
    \toprule
    Name$^a$ & $\chi^2/dof$ & log($Z$) & Parameters \\
    \midrule
    \multicolumn{4}{c}{\textit{One-Component Models}} \\
    \midrule
         \textsc{diskpbb}$^b$ & 475.22/431 & $-$258.1 & $k_{\rm B}T_{\rm in}= 117^{+1}_{-2}$eV \\
         & & & $p = 0.34\pm 0.01$ \\
         & & & \\
         \textsc{comptt}$^c$ & 495.51/430 & $-266.4$ & $k_{\rm B}T_{\rm 0} = 49\pm1$ eV\\
         & & & $k_{\rm B}T_{\rm e} < 50$ keV\\
         & & & $\tau < 1.2$ \\
         & & & \\
         \textsc{nthcomp}$^d$ & 469.79/431 & $-251.6$ & $k_{\rm B}T_{\rm 0} = 62\pm 2$ eV\\
         & & & $k_{\rm B}$ \textit{unconstrained} \\
         & & & $\Gamma=6.09\pm0.15$\\
         & & & \\
         \textsc{zlogpar}$^e$ & 453.48/431 & $-239.7$ & $a = 2.74\pm0.18$\\
         & & & $b = 2.73\pm0.21$ \\
         & & & $E_{\rm p}=0.2$ keV (\textit{fixed}) \\
    \midrule
    \multicolumn{4}{c}{\textit{Two-Component Models}} \\
    \midrule
    \textsc{zbbody+} & 459.17/430 & $-246.1$ & $k_{\rm B}T_1 = 53 \pm 2$ eV \\
    \textsc{zbbody}$^f$ & & & $k_{\rm B}T_2 = 109^{+4}_{-3}$ eV \\
    & & & \\
    \textsc{zbbody+} & 455.47/428 & $-239.2$ & $k_{\rm B}T_\textrm{BB} = 51^{+3}_{-4}$ eV \\
    \textsc{comptt} & & & $k_{\rm B}T_{\rm e} < 31$ keV \\
    & & & $\tau < 0.02$ \\
    & & & \\
    \textsc{tbabs$*$} & 466.52/428 & $-250.0$ & $\tau = 5.40^{+0.26}_{-0.13}$ \\
    \textsc{thcomp(zbb)}$^g$ & & & $k_{\rm B}T_{\rm e} < 1.2$ keV\\
    & & & $T_{\rm 0} = 53\pm2$ eV \\
    & & & $c_f > 0.95$ \\
    & & & \\
    \textsc{comptt} & 454.46/427 & $-258.0$ & $k_{\rm B}T_{0,1} = 45\pm2$~eV \\
    \textsc{comptt} & & & $T_{\rm e,1} = 27^{+13}_{-7}$~keV \\
    & & & $\tau_1<$0.3 \\
    & & & $k_{\rm B}T_{0,2} = 97\pm1$ \\
    & & & $T_{\rm e,2} = 6^{+7}_{-3}$ \\
    & & & $\tau_2<0.05$ \\
    \bottomrule
    \end{tabular}
    \flushleft{\scriptsize{
    a) All models include a \textsc{tbabs} term to account for Galactic absorption.\\
    b) \textsc{diskpbb} models a disk emission profile where the disk surface-temperature profile $T \propto r^{-p}$ and the disk has an inner edge with temperature $T_\textrm{in}$.\\
    c) \textsc{comptt} \citep{TITARCHUCK_1994} represents a Comptonised emission component with parameters seed photon temperature $T_{\rm 0}$, plasma electron temperature $k_{\rm B}T_{\rm e}$, and optical depth $\tau$.\\
    d) \textsc{nthcomp} models a multi-colour blackbody Comptonised into a power law \citep{ZDZIARSKI_1996,ZYCKI_1999}, set by $T_{\rm 0}$, $k_{\rm B}T_{\rm e}$, and power-law photon index $\Gamma$.\\
    e) \textsc{zlogpar} is purely phenomenological and is set by $A(E) = K (E(1+z)/E_{\rm p})^{(-a-b{\rm log}( E(1+z)/E_{\rm p}))}$.\\
    f) \textsc{zbbody} is the standard redshifted blackbody, set by $k_{\rm B}T$.\\
    g) \textsc{thcomp} \citep{ZDZIARSKI_2020} yields a modified blackbody, based on parameters $\tau$, $k_{\rm B}T_{\rm e}$, $k_{\rm B}T_{\rm 0}$, and covering factor $c_f$.\\
    }}
    \label{tab:xray_xmmfit}
\end{table}

\vspace{-0.3cm}
\subsection{\textit{Swift}-XRT analysis}
For the \textit{Swift}-XRT observations we tested several models in \textsc{Xspec} using Cash statistics (Section~\ref{sec:xray_swift}). We tested a single power-law, a single blackbody, and a double blackbody (2BB), the summary results of which are included in Table~\ref{tab:xray_lc}. We present a full overview of the results of our fitting of the \textit{Swift}-XRT spectra in Table~\ref{tab:xrtfits}.

\begin{table*}[!h]
\renewcommand*{\arraystretch}{1.4}
    \centering
    \caption{
    textit{Swift}-XRT spectral fitting results.
    }
    \begin{tabular}{l|ccc|ccc|cc} 
    \toprule
        & \multicolumn{3}{c}{Single Power Law} &  \multicolumn{3}{c}{Single blackbody} &  \multicolumn{2}{c}{Dual blackbody$^a$} \\
        & $\Gamma$            &  $A_1^b$ $\times$ $10^{-4}$   & $C/dof$ &   $k_{\rm B}T$ &  Norm.    &  $C/dof$  &  Norm. &   $C/dof$  \\
    Obs &    &  (ph cm$^{-2}$ s$^{-1}$ keV$^{-1}$) &      &  (eV)  & ($10^{-4}$) &      &  ($10^{-4}$)   &  \\
    \midrule
    20-12-2020  & $4.0\pm0.8$         & $5.5^{+4.3}_{-2.6}$ & 34.07/47  & $93^{+17}_{-15}$ & $1.16^{+0.55}_{-0.32}$ & 32.65/47 & $2.87^{+0.56}_{-0.53}$ & 34.94/48 \\
    27-12-2020  & $3.9\pm0.7$         & $7.2^{+4.6}_{-3.2}$ & 38.65/47  & $94^{+14}_{-12}$ & $1.40^{+0.46}_{-0.32}$ & 40.80/47 & $3.44^{+0.57}_{-0.54}$ & 48.40/48 \\
    06-01-2021  & $4.4^{+1.1}_{-0.9}$ & $4.2^{+4.3}_{-2.6}$ & 40.38/47  & $79^{+20}_{-18}$ & $1.70^{+1.62}_{-0.63}$ & 41.40/47 & $3.09^{+0.69}_{-0.62}$ & 40.40/48 \\
    18-01-2021  & $4.5*$              & $1.2\pm0.7$         & 41.76/47  & 82*             & $0.52^{+0.28}_{-0.26}$ & 48.47/47 & $0.98^{+0.55}_{-0.51}$ & 49.20/48 \\
    22-01-2021  & $5.6^{+1.9}_{-1.4}$ & $0.8^{+2.0}_{-0.7}$ & 48.22/47  & $63^{+18}_{-17}$ & $1.56^{+2.83}_{-0.78}$ & 45.05/47 & $1.57^{+0.45}_{-0.42}$ & 49.28/48 \\
    \bottomrule
    \end{tabular}
    \flushleft{\scriptsize{
    a) In the dual-blackbody fit, temperatures were kept frozen at 54 eV and 109 eV with the normalisation of the higher-temperature blackbody set to 0.13 times that of the lower-temperature blackbody, matching the \textit{XMM-Newton} observation (see Section~\ref{sec:xray_swift}). The table lists the normalisation of the lower-temperature blackbody for this dual-blackbody scenario.\\
    b) $A_1$ is the power-law normalisation at 1 keV.\\
    }}
\label{tab:xrtfits}
\end{table*}

\vspace{-0.3cm}
\subsection{NICER reductions and analysis}
\label{sec:app_xray_nicer}
To maximise the observing time before the Sun-block period, the observations were pushed as close as possible to the instrumental limits of \textit{NICER}. These limits are set by a maximum X-ray contamination by optical loading. As the effect of the optical loading was larger than expected, a significant fraction of the data unfortunately needed to be rejected, particularly close to the Sun-block period. The final good time intervals (GTIs) cover the period 6 to 11 January (MJD 58854--58859) and represent a combined exposure time of 18.5 ks.

Observations are separated into time intervals (TIs) with exposures between 30-60 sec for filtering, and a background spectrum is constructed for each TI using the 3C50 background estimator. Absolute count rates from 50 Focal Plane Modules (FPMs) are used, excluding noisy detectors 14 and 34. Two indicators of optical loading are chosen to filter the dataset: residual background and undershoot rate. The first, based on signal in an energy band that has negligible source contribution, follows the level 3 filtering guidelines standardised in \citet[][]{REMILLARD_2022}. This filtering process excludes TIs in which 3C50 underestimates the background due to elevated levels of optical loading (0.2–0.3 keV) or the high energy particle background (13–15 keV), indicating an unreliable source spectrum for the TI. Periods of underestimated optical loading are identified as those intervals where the background-subtracted count rate in the 0.2--0.3 keV band is greater than 2 photons per second. For periods with count rates below this limit, the optical loading was found to be sufficiently low to be able to accurately model the background for all energy bands. For periods with a 0.2--0.3 keV count rate above the limit, the optical loading proved too strong to accurately model the background at energies above 0.3 keV, and these periods were therefore excluded from our GTIs. 

Second, TIs affected by optical loading at energies above 0.3 keV are screened out using the detector undershoot rate. This is an instrumental parameter that counts detector resets due to the accumulation of electrons in the equipment freed by incident optical photons \citep[][]{REMILLARD_2022}. Optical loading and the associated undershoot rates are highest at low Sun angles, reaching a maximum of 300-400 counts~s$^{-1}$ closest to the Sun-block period. For TIs with an undershoot rate above 150 counts~s$^{-1}$, a correlated relationship between undershoot rate and count rate in the 0.3–2 keV source sensitivity band is observed. This indicates that in-band count rates are significantly contaminated by optical loading when the undershoot rate is greater than 150 counts~s$^{-1}$, even when the contribution from optical loading is modelled appropriately in the 0.2-0.3 keV band. Because 3C50 currently lacks the capability to model the effect of extreme levels of optical loading at energies above 0.3 keV, we excluded observations with an undershoot rate $>$150 counts~s$^{-1}$ from our GTIs. 

For periods where both background residuals and undershoot rate are below their respective limits, the optical loading was found to be sufficiently low to be able to accurately model the background for all energy bands. We subsequently use the \textit{NICER} reduction pipeline to produce the standard RMF and ARF. We fit the \textit{NICER} data using \textsc{Xspec} v12.12.0, which is part of HEASOFT v6.28. The data have been re-binned to a minimum of 25 counts per bin, and we use $\chi^2$-statistics.

The \textit{NICER} spectrum of J234402 shows soft emission. As the count rate above 1.2~keV is relatively low, we experimented with different fitting ranges to assess the impact of the presence of noise in this energy range. Performing all fits described above in the energy ranges 0.3--2.0~keV and 0.3--1.2~keV, we find that 2BB (Section~\ref{sec:xray_xmm}) is the best-fitting model in both cases, with no significant impact on the goodness-of-fit. In fact, the only significant impact we find is that the power-law+blackbody proves a slightly better model in the 0.3--1.2 keV range with $\chi^2/dof$ increasing from 179.3/134 (in the 0.3--2.0~keV range) to 103.6/84 (in the 0.3--1.2~keV range). We interpret this to mean that the power-law component, evidently a poor fit to our data, is better constrained when using the broader energy range due to the lack of signal above 1.2 keV. We therefore chose to use the 0.3--2.0~keV range for fitting the \textit{NICER} data, as it provides the best constraints for our models, while avoiding the background-dominated energy range above 2 keV. The model fitting results quoted for \textit{NICER} have all been derived in the 0.3--2.0 keV range.

To calculate the \textit{NICER} fluxes included in Figure~\ref{fig:xray_combined}, we scaled the 0.3--2.0 count rate using a 0.2--2.0~keV flux based on our best-fit 2BB model, which was fit to the combined \textit{NICER} data. Although there is some spectral variability, we find that the effect of the spectral changes on the count-to-flux scaling is negligible.

\section{UV and optical observations}
\label{sec:app_optuv}

\subsection{Photometry}
\begin{figure*}[!h]
    \centering
    \includegraphics[width=0.5\linewidth]{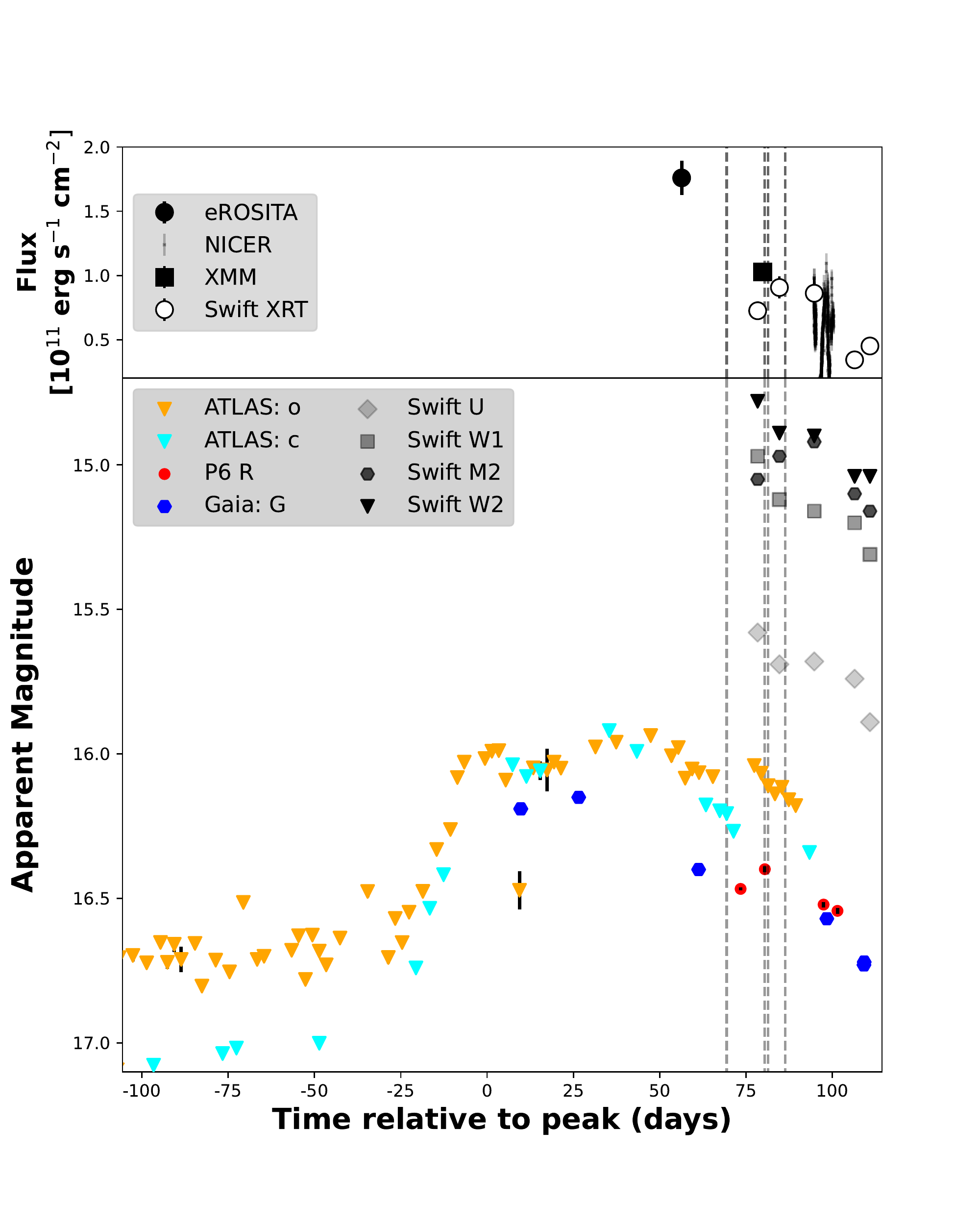}
    \caption{The optical, UV, and X-ray light curves around the time of ignition. Fluxes and magnitudes are corrected for Galactic absorption. $Top$: The X-ray light curve, consisting of the eROSITA, \textit{XMM-Newton}, \textit{Swift}-XRT, and \textit{NICER} data. $Bottom$: Light curves for ATLAS $c$ and $o$, \textit{Gaia} $g$, \textit{Swift}-UVOT, and PROMPT6 R data. These data are not host subtracted. The data are plotted in time relative to the point of the first peak in the $o$-band light curve (see Section~\ref{sec:mw_opt_lc}). The dashed vertical lines represent the epochs for our optical spectroscopy. The optical data show a sharp increase over approximately 3 weeks, followed by a sharp turnover, a further bump visible in the $o$ and $c$ bands, and subsequent decline. The decline in optical brightness is matched in the UV and X-ray data.}
    \label{fig:opt_lc}
\end{figure*}

In Figure~\ref{fig:opt_lc} we show a close-up of our photometric light curve (not host-corrected) in combination with our UV and X-ray data. The figure also shows the epochs for our optical spectroscopy.

\subsubsection{POSS2 photometry}
J234402 was observed with the UK Schmidt telescope in the POSS2 B band in 1985 (MJD 46267) and the POSS2 R band in 1997 (MJD 50691). The blend of the three southern sources was identified as a single target for the 2dFGRS \citep{COLLESS_2003}. The extraction of the POSS2 photographic flux values is complex, as it depends on the responses of the  emulsion + filter combinations used in the POSS2 observations. We make use of the magnitudes calculated for the SuperCOSMOS Survey \citep{HAMBLY_2001}. In the POSS2-R image, magnitudes were extracted for both the northern target and the blended group of southern targets. However, for the POSS2-B imaging, a magnitude is only available for the combined flux from the cluster consisting of all four objects. 

To estimate the impact of the blending in the POSS2-B image, we downloaded the reduced image from the MAST DSS server. We use Astropy \citep[][]{ASTROPY_2013,ASTROPY_2018} routines to create a background map for the image, automatically detect sources in the background-subtracted data, and separate these sources where possible. The routine successfully separates J234402 from the blend of the three southern sources. We find that approximately 40$\%$ of the combined flux is associated with J234402. Adjusting the SuperCOSMOS magnitude by this fraction, we find that the POSS2-B magnitude for J234402 is 18.2 and POSS2-R magnitude is 16.9. \citet{HAMBLY_2001} quote the average accuracy of the POSS photometry as within 0.3 mag. We note that since the POSS2 photometric system differs from that of our more recent observations, a direct comparison is difficult. However, these data remain valuable as a constraint on the longer term evolution of J234402. 

\subsubsection{\textit{Swift}-UVOT photometry}
\label{sec:app_optuv_swift}
\begin{figure*}[!h]
    \centering
    \includegraphics[width=0.5\linewidth]{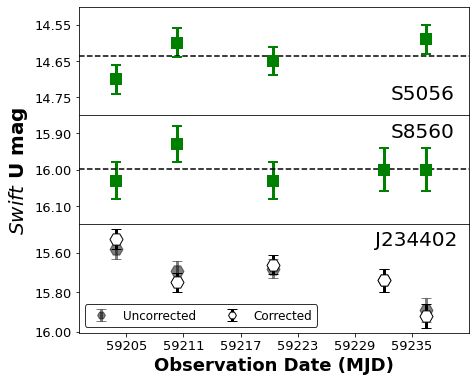}
    \caption{Example of adjusting \textit{Swift}-UVOT fluxes, for the U-band. The magnitudes of two standard stars are shown in the top two panels (the magnitude of the first standard star could not be extracted for the fourth epoch). We assume the flux of these objects to be constant over the time covered by the \textit{Swift} observations. We calculate the difference with the mean flux (indicated by the dashed line) for the observed flux in each epoch and adjust the magnitudes for J234402 accordingly. The bottom panel shows the uncorrected (grey) and corrected (open) magnitudes for the U-band observations of J234402.} 
    \label{fig:UV_swift_cal}
\end{figure*}

We test for the contribution of instrumental variability to the flux changes among observations. We correct the UVOT flux measurements by calibrating to two stars in the UVOT field of view (source IDs 2311717768560445056 and 2311905170869138560, referred to here as S5056 and S8560, respectively). We illustrate this method in Figure~\ref{fig:UV_swift_cal}. For the U, W1, and W2 filters, we calculate the average flux for the standard stars. In each filter, we calculate the difference between the observed and the average flux per epoch. We assume that these deviations are not intrinsic to the sources, bur rather are instrumental, and we adjust the magnitudes for J234422 accordingly. For M2, we have insufficient epochs with a good detection of the reference stars, so we cannot correct the flux in this filter. This is also the reason that we cannot apply this method to the \textit{XMM-Newton} OM-UVM2 data to correct for inter-mission uncertainties. The uncorrected and corrected magnitudes are included in Figure~\ref{fig:UV_swift_cal}, to show the size of the correction.

\subsection{Optical spectroscopy}
\subsubsection{Location of the ignition event}
\begin{figure*}[!h]
    \centering
    \begin{minipage}{0.34\textwidth}
    \vspace{-.7cm}
        \includegraphics[width=1.1\textwidth,angle=-90]{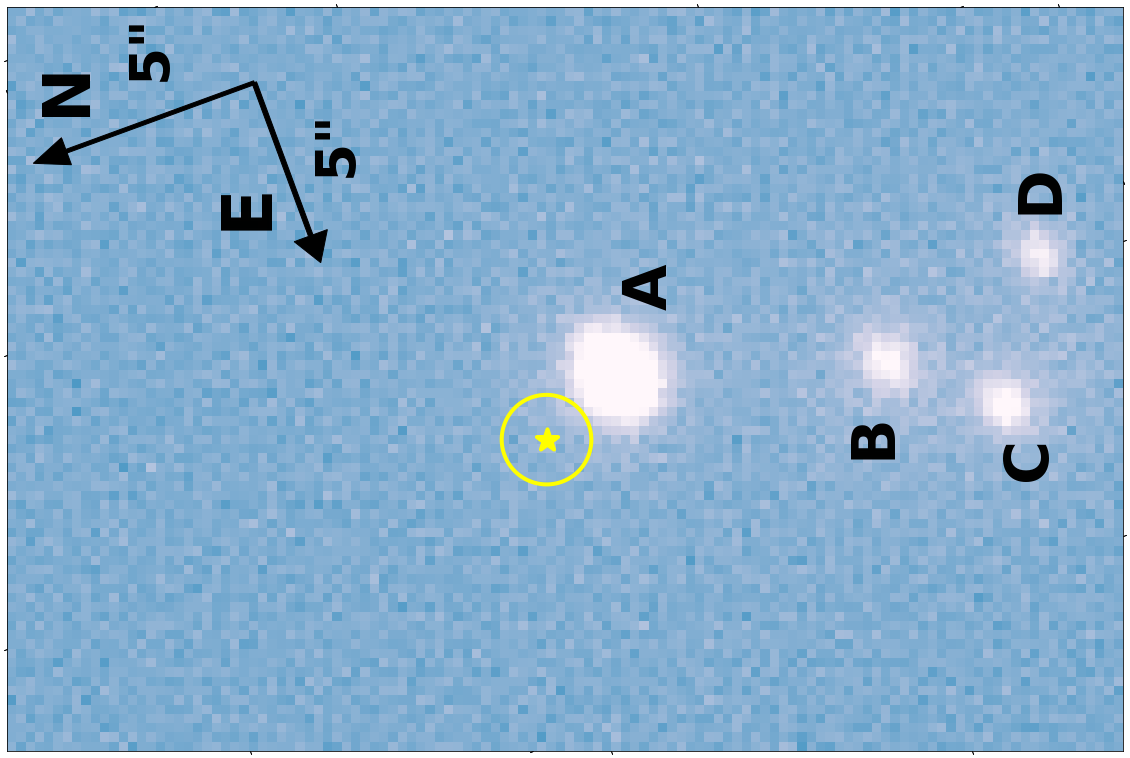}
    \end{minipage}
    \begin{minipage}{0.65\textwidth}
        \hspace{-1.7cm}
        \includegraphics[trim=4.6cm 0.2cm 3.0cm 2.2cm,clip,width=1.1\textwidth]{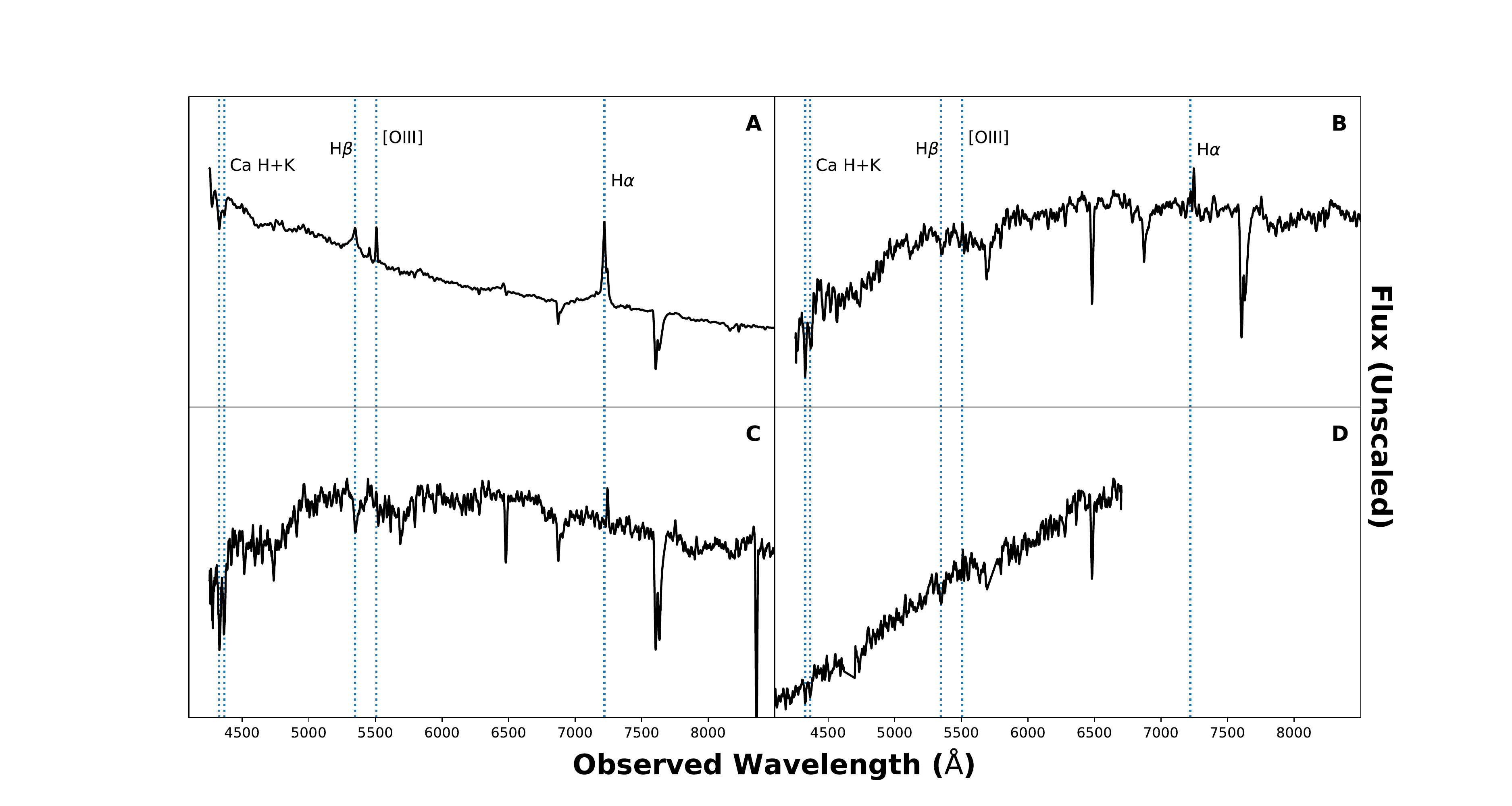}
    \end{minipage}
    \caption{Counterpart and environment of J234402. \textit{Left}: The optical counterpart of J234402. The yellow star represents the X-ray position and the yellow circle indicates the 1.0" (1$\sigma$ radius) positional uncertainty of the eROSITA detection. This image was taken as an acquisition exposure with the Baade telescope using a white filter during our spectroscopic follow-up campaign. The source area includes four targets (labelled A--D), which were found to be at the same redshift. The Northern object, labelled A, is associated with the significant X-ray-flux increase detected by eROSITA, as well as with the significant optical outburst seen with Gaia (TNS\#85552).
    \textit{Right}: Overview of the spectra taken for the four objects that make up the galaxy group containing J234402. The labels A through D correspond to the objects in the image on the left. Object A is J234402. The spectra are plotted at their observed wavelengths, and the positions of several spectral lines, at $z=0.100$ are shown in each plot. The plots illustrate the clear difference in spectral classification among the objects --- object A shows an AGN-like spectrum, whereas the other three objects appear to be quiescent galaxies. The flux scale is arbitrary, and we note that the spectrum for object D has not been flux-calibrated. Spectra for A, B, and C are from the Baade (Magellan) observation. The spectrum for D is from the SALT observation of MJD 59206 (23 December).}
    \label{fig:spec_group}
\end{figure*}

Our follow-up optical spectroscopy shows that the northern galaxy (in the group of four) has a strong blue continuum, as well as broad Balmer emission lines. Spectra for the southern objects in the group resemble those of quiescent galaxies. This agrees with the eROSITA and Gaia localisations of the brightening X-ray and optical sources, to identify the northern galaxy as the location of the ignition event. Spectra for all four objects are shown in Figure~\ref{fig:spec_group}. The three southern objects are at approximately the same redshift as the bluer northern source with \textit{z} ranging from 0.099 to 0.101. The match in redshift indicates the four objects are likely in physical proximity and form a small group. Using the angular separation between objects A and D, approximately 9", as a rough estimate for the maximum extent of the group, we find a transverse proper distance of approximately 18~kpc. 
\subsubsection{The FORS2 spectrum}
The FORS2 spectrum shows several distinct features at wavelengths lower than 4000 \AA (Figure~\ref{fig:opt_spec}). The absence of these features in the near-contemporaneous SALT spectrum (23 Dec.) led us to further investigate the `bumps' in the FORS2 spectrum. We checked the flux calibration by comparing the flux-scaled spectrum of the standard star (HZ4) with the tabled archival data for that star. We found that in the wavelength range below 4000 \AA\ the flux-calibrated spectrum is overestimated in a pattern that matches the `bumps' seen in the spectrum of J234402. The discrepancy between the flux-scaled, observed spectrum and the tabled data in this region of the spectrum is as large as 10\%, compared to $<$3\% in the rest of the spectrum.

\subsubsection{Fitting a blackbody temperature}
\begin{figure*}[!h]
    \centering
    \includegraphics[width=0.8\linewidth]{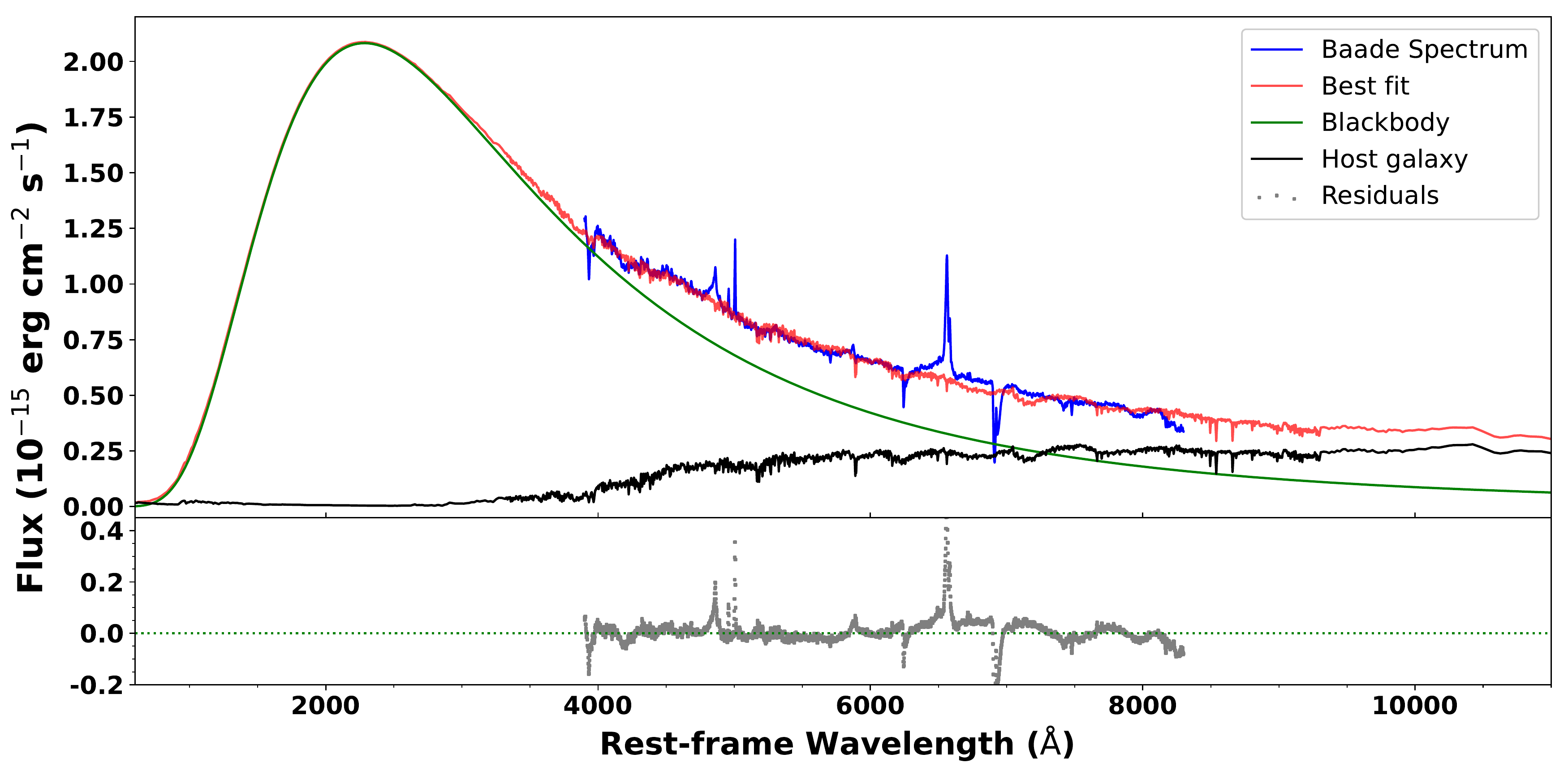}
    \caption{Illustration of our fitting procedure for optical spectra with blackbody+host model. The example spectrum was taken with the Baade telescope on MJD 59195. The top panel shows the data and model, and the bottom panel shows the residuals. The fitting model consists of the Planck function (\textit{green}) and a template for the emission from the host galaxy (\textit{black}). The best-fit model (\textit{red}) is defined by log(T$_\mathrm{bb}$/[K])~=~4.10 and results in a bolometric luminosity log(L$_\mathrm{bb}$/[erg s$^{-1}$])~=~44.3.}
    \label{fig:opt_specfit_bb}
\end{figure*}

We illustrate our fitting procedure for a blackbody spectrum to the optical spectrum, in Figure~\ref{fig:opt_specfit_bb}. The procedure is similar to that described in Section~\ref{sec:mw_spec_fit}. We include a host-galaxy template and a blackbody spectrum. The parameters of the host-galaxy template are fixed to the values found for the Baade spectrum. We let the blackbody temperature and a normalisation parameter vary freely.

\end{appendix}

\end{document}